\documentclass{ajour}
\usepackage{epsfig}

\renewcommand{\arraystretch}{1.3}

\newcommand{\equ}{Eq.}
\newcommand{\equs}{Eqs.}
\newcommand{\fig}{Fig.}
\newcommand{\figs}{Figs.}
\newcommand{\sect}{Sec.}
\newcommand{\rem}[1]{}

\newcommand{\FIGo}[3]{\begin{figure}%
#3%
\caption[]{\footnotesize #2}%
\label{#1}%
\end{figure}}

\rem{
\newcommand{\FIGo}[3]{%
\begin{figure}{\caption{}\label{#1}}\end{figure} \vspace*{1.5ex}
}
\newcommand{\printfigcap}[2]{\rule{0cm}{1cm}Figure~\ref{#1}: #2 \\}
\newcommand{\showfig}[1]{\begin{figure} #1
  \vbox{\vspace*{2.0cm}} \caption{Jan Wiersig} \end{figure} \clearpage }
}

\newcommand{\stdarray}{\renewcommand{\arraystretch}{1.1}}

\newcommand{\Eell}{{\cal E}}

\newcommand{\Kell}{{\cal K}}

\newcommand{\const}{\mbox{const}}

\newcommand{\maslov}{\alpha}

\newcommand{\maslovd}{\tilde{\alpha}}

\newcommand{\h}{E}
\newcommand{\bm}[1]{\mbox{\boldmath$#1$\unboldmath}}
\newcommand{\Bomega}{\bm \omega}
\newcommand{\BI}{\bm I}

\newcommand{\dof}{N}
\newcommand{\scp}[2]{#1 \cdot #2}
\newcommand{\scpi}[1]{#1^2}

\newcommand{\qcf}{\rho}

\newcommand{\Ident}[1]{{\mbox{{\bf 1}}}_{#1}}

\newcommand{\phia}{\varphi}
\newcommand{\Ia}{J}
\newcommand{\Bphia}{{\bm \varphi}}
\newcommand{\BIa}{{\bm J}}

\newcommand{\phib}{\vartheta}
\newcommand{\Ib}{P}
\newcommand{\Bphib}{{\bm \vartheta}}
\newcommand{\BIb}{{\bm P}}

\newcommand{\Ic}{I}
\newcommand{\BIc}{{\bm I}}

\newcommand{\Icred}{\tilde{\Ic}}
\newcommand{\BIcred}{\bm {\tilde{I}}}

\newcommand{\Id}{L}
\newcommand{\BId}{{\bm L}}

\newcommand{\patch}{\Lambda}
\newcommand{\M}{M}
\newcommand{\eps}{\varepsilon}
\newcommand{\mmatrix}{Q}
\newcommand{\mamatrix}{Q_2}
\newcommand{\mbmatrix}{W}
\newcommand{\nullm}{\mbox{{\bf 0}}}

\newcommand{\Es}{E_\dof}
\newcommand{\sign}{{\rm sign}}
\newcommand{\dio}{d}
\newcommand{\Nindex}{\Sigma}
\newcommand{\Kindex}{\Sigma}
\newcommand{\stIb}{\bm L}
\newcommand{\m}{\bm m}
\newcommand{\Happ}{H_{\mbox{app}}}

\newcommand{\na}{k}
\newcommand{\Bna}{{\bm k}}
\newcommand{\nb}{n}
\newcommand{\Bnb}{{\bm n}}
\newcommand{\maslova}{\beta}
\newcommand{\Bmaslova}{{\bm \beta}}
\newcommand{\maslovb}{\alpha}
\newcommand{\Bmaslovb}{{\bm \alpha}}
\newcommand{\EFP}[1]{\Omega(#1,\nb_\dof)}

\newcommand{\EFpsib}[1]{{\psi}(#1,\nb_\dof)}
\newcommand{\EFpsic}[1]{\Psi(#1,\nb_\dof)}

\begin{document}

\authorrunninghead{Wiersig}
\titlerunninghead{Resonance zones in action space}


\title{Resonance zones in action space}
\author{Jan Wiersig}
\affil{Max-Planck-Institut f\"ur Physik komplexer Systeme, D-01187 Dresden,
Germany} 
\date{\today}
\email{jwiersig@mpipks-dresden.mpg.de}

\abstract{
The classical and quantum mechanics of isolated, nonlinear resonances in
integrable systems with $\dof \geq 2$ degrees of freedom is discussed in terms
of geometry in the space of action variables.   
Energy surfaces and frequencies are calculated and graphically presented for
invariant tori inside and outside the resonance zone.
The quantum mechanical eigenvalues, computed in the semiclassical WKB
approximation, show a regular pattern when transformed into the action space
of the associated symmetry reduced system: eigenvalues inside the resonance
zone are arranged on $\dof$-dimensional cubic lattices, whereas those outside
are, in general, non-periodically distributed. However, $\dof$-dimensional
triclinic (skewed) lattices exist locally. Both kinds of lattices are joined
smoothly across the classical separatrix surface.    
The statements are illustrated with the help of two and three coupled rotors. 
}


\rem{
\vfill\noindent 65 pages, 19 Figures\clearpage
{\large\noindent\vspace*{3ex}\\[3ex]
Jan Wiersig\\[1.5ex]
Max-Planck-Institut f\"ur Physik komplexer Systeme\\[1ex]
D-01187 Dresden, Germany\\[1ex]
\begin{tabbing}
phone: \= \kill 
phone: \> +49 351 871 1223\\
fax:   \> +49 351 871 1999\\
email: \> jwiersig@mpipks-dresden.mpg.de\\
\end{tabbing}
}
\clearpage
}

\begin{article}

\section{INTRODUCTION}
\label{sec:intro}
The surfaces of constant energy $H(\BIc) = \h$ in the space of action
variables $\BIc = (\Ic_1,\ldots,\Ic_\dof)$ contain the essential information
about the dynamics of a compact integrable system, such as the fundamental
frequencies ${\bm \omega} = \partial H/\partial \BIc$ and the foliation by
invariant tori.
The frequencies are given by the normals of the surfaces, whereas the foliation
can be read off from the way an energy surface is divided into several patches,
each representing a certain type of motion. 
``Simple systems'' with trivial foliation like uncoupled harmonic oscillators
have globally continuous and smooth energy surfaces. For many of the non-simple
systems considered so far 
\cite{Richter90,DJR94,DW94,RW95,Heudecker95,WR96,RDWW96,DHJPSWWW97,Wiersig00b},
the energy surfaces are nonsmooth at separatrices, but, nevertheless,
continuous or can be made continuous by a reduction of the system's discrete
symmetries. In~\cite{Wiersig98,WWD98,Wiersig00b} such systems are referred to
as ``one-component systems''.          

The subject of the present work is a class of one-component systems composed of
integrable approximations of near-integrable systems of the form      
\begin{equation}\label{eq:ham0}
  H(\BIa,\Bphia) = H_0(\BIa)+\eps\, H_1(\BIa,\Bphia) \ ,
\end{equation} 
where $\eps > 0$ is the perturbation parameter, $H_1$ is a periodic
function in the angles $\Bphia$, and $H_0$ describes the unperturbed system
which is required to be simple in order to ensure that its action-angle
variables $\BIa,\Bphia$ can be defined globally in phase space. 
The KAM theorem~\cite{Kolmog57,Arnold63,Moser62} states that a
finite fraction of the unperturbed invariant tori survives smoothly deformed,
namely those tori ``sufficiently far'' from resonance surfaces    
\begin{equation}\label{eq:resonanceplane}
\scp{\bm m}{{\bm \omega}_0 (\BIa)} = 0 \ ,
\end{equation}
with relatively prime integer vectors $\m$ (an entire set $m_1,\ldots,m_\dof$
has no common divisor).   
The remaining fraction consists of chaotic trajectories, primary ``islands''
associated with the resonances of the unperturbed
system~(\ref{eq:resonanceplane}) and higher-order islands.   
The smoothly deformed tori can be approximated by canonical perturbation
theory. In the first order, the Hamiltonian~(\ref{eq:ham0}) is averaged over
the unperturbed tori, implicitly assuming that all angles $\Bphia$ are rapidly
varying.   
Close to a primary island this assumption is not valid; the phase $\scp{\bm
m}{\Bphia}$ is almost stationary. 
Resonant perturbation theory~\cite{Chir79} then suggests a better 
procedure: average over submanifolds of the unperturbed tori, $\scp{\bm
m}{\Bphia} = \const$; expand the resulting integrable Hamiltonian
\begin{equation}\label{eq:hama}
  H(\BIa,\Bphia) = H_0(\BIa)+\eps\, V(\BIa,\scp{\bm m}{\Bphia})
\end{equation} 
at the resonance surface in $\BIa$ and in a Fourier
series in $\Bphia$, and retain only the most important contribution
\begin{equation}\label{eq:potential}
  V(\BIa,\scp{\bm m}{\Bphia}) = f(\BIa)\cos{(q\scp{\bm m}{\Bphia})}
 \ ,
\end{equation} 
where $q$ is an integer.
The regular dynamics in- and outside an isolated primary island described by the integrable
Hamiltonian~(\ref{eq:hama})-(\ref{eq:potential}) is well understood; see,
e.g., \cite{LichLieb92}. Hamiltonians of this type have been frequently used
as physical models, e.g., for energy transfer in triatomic molecules;
see~\cite{Sibert82b} and references therein.    
However, energy surfaces have only been presented for a special case with two
degrees of freedom~\cite{DW94}. We here show the energy surfaces for a broader
class of isolated-island systems with $\dof \geq 2$ degrees of freedom.  

Actions were the central ingredients for the old quantum mechanics before
1926. Bohr and Sommerfeld, among others, computed energy spectra by 
discretizing classical action integrals in integer multiples of $\hbar$,
Planck's constant divided by $2\pi$.   
The necessity of classical integrability was pointed out by Einstein, who
formulated the quantization rules in terms of invariant tori and action
variables~\cite{Einstein17}.     
Later, Brillouin derived from Schr{\"o}dinger's equation that the 
quantization of actions is an approximation rigorously valid only in the
(semi-)classical limit $\hbar \to 0$.     
Keller finally corrected this semiclassical approximation by Maslov
indices in the presence of caustics~\cite{Keller58}.   
According to the Einstein-Brillouin-Keller (EBK) rule, the
quantization of an integrable system is a discretization of its action
space by a $\dof$-dimensional cubic lattice with lattice constant $\hbar$.  
Unfortunately, this plain recipe only applies to simple systems because the
presence of separatrices destroys the applicability of the EBK rule, and, more
severely, there is in general no one-to-one correspondence between classical
action variables and quantum eigenvalues~\cite{Wiersig98}. However, the recipe
can be extended to one-component systems by introducing the action space of the
associated symmetry reduced system~\cite{Wiersig98}.  
It has been found in~\cite{RDWW96,WWD97,WWD98} that the classical partition of
this space into domains of different types of motion carries
over to the discretization: away from the separatrix surfaces there exist
$\dof$-dimensional cubic lattices, each related to the EBK rule for
the corresponding type of motion; across the separatrix surfaces the lattices
are smoothly connected.      
We here demonstrate that the Hamiltonian~(\ref{eq:hama})-(\ref{eq:potential})
gives rise to novel, less symmetric eigenvalue lattices.

The outline of the paper is as follows. 
In \sect~\ref{sec:classic}, we transform the
Hamiltonian~(\ref{eq:hama})-(\ref{eq:potential}) to a simpler form and expand
it at the resonance surface in accordance to resonant perturbation theory. We
then compute action variables and energy surfaces. The latter are
graphically presented for a simple model, two and three coupled rotors. The
section ends with a comparison to canonical perturbation theory.  
In \sect~\ref{sec:quantum}, we firstly derive the semiclassical quantization
condition, and then discuss the eigenvalue pattern in action space,
illustrated with the help of the coupled-rotors model.
Finally, we briefly draw conclusions in \sect~\ref{sec:con}.  

\section{ENERGY SURFACES IN ACTION SPACE}	
\label{sec:classic}
\subsection{TRANSFORMATION TO STANDARD FORM}
First of all, we transform the Hamiltonian~(\ref{eq:hama}) such that $\m
\rightarrow (0,\ldots,0,1)$. If only one component of $\m$ is nonzero then
this is achieved by redefining the indices. In the general case we first
redefine the indices such that the first two components $m_1,m_2$ are nonzero 
and relatively prime. 
Second, we introduce new phase space variables $(\BIb,\Bphib)$ with one of
the new angles $\phib_\dof = \scp{\bm m}{\Bphia}$ being stationary at the
resonance surface. For this purpose, we apply the generating function
$F(\Bphia,\BIb) = \scp{\BIb}{\mmatrix \Bphia}$ of Goldstein type
2~\cite{Gold80}. The relations $\Bphib = \partial F/\partial \BIb$ and $\BIa =
\partial F/\partial \Bphia$ give  
\begin{equation}\label{eq:trans1}
  \BIb = (\mmatrix^t)^{-1} \BIa \ , \quad \Bphib = \mmatrix \Bphia \ .
\end{equation}
This transformation is not only canonical but also unimodular provided that 
the $\dof\times\dof$-matrix $\mmatrix$ has integer-valued components and 
determinant $\pm 1$ (For our purpose it is sufficient to consider matrices
with determinant $1$). This property ensures that $(\BIb,\Bphib)$ are
action-angle variables of the unperturbed system. 
Remarkably, non-unimodular transformations are often used in the literature,
even though the resulting variables are not action-angle variables in the
sense of Liouville-Arnol$'$d~\cite{Arnold78}, i.e. fixing the actions and
varying the angles independently from $0$ to $2\pi$ does not yield a single
complete cover of a torus. This has already been mentioned in~\cite{RobLit87}
where a construction of the matrix $\mmatrix$ has been given. We here make up
a similar, but simpler matrix. For two-degrees-of-freedom systems we simply
choose  
\begin{equation}\label{eq:mamatrix}
  \mamatrix = \left(
  \begin{array}{cc}
        \dio_2 & -\dio_1 \\
        m_1 & m_2  \\
  \end{array}
  \right)
\end{equation}
where $\dio_1,\dio_2$ are integers satisfying the diophantine
equation    
\begin{equation}\label{eq:dio}
        \det{\mamatrix} = \dio_1 m_1 + \dio_2 m_2 = 1 \ .
\end{equation}
It is known from number theory (see, e.g., \cite{Rieger76}) that for given
$m_1,m_2$, there exist a fundamental solution $\dio_1,\dio_2$, which can be
calculated by Euclid's algorithm or chosen by hand. From this one gets a whole
series of solutions $\dio_1(n) = \dio_1 - m_2\,n$, $\dio_2(n) = \dio_2 +
m_1\,n$ with an integer $n$.
With the matrix $\mamatrix$, the $(\dof-2)\times (\dof-2)$ unit matrix
$\Ident{\dof-2}$ and the $(\dof-2)\times 2$ matrix  
\begin{equation}\label{eq:mbmatrix}
  \mbmatrix = \left(
  \begin{array}{ccc}
        0 & \ldots & 0 \\
        m_3 & \ldots & m_\dof\\
  \end{array}
  \right)
\end{equation}
we construct for more than two degrees of freedom the $\dof\times\dof$ matrix
\begin{equation}\label{eq:mmatrix}
  \mmatrix = \left(
  \begin{array}{cc}
        \nullm & \Ident{\dof-2} \\
        \mamatrix & \mbmatrix \\
  \end{array}
  \right) \ .
\end{equation}
Using this matrix in transformation~(\ref{eq:trans1}),
Hamilton's function~(\ref{eq:hama}) takes the new form
\begin{equation}\label{eq:hamb}
  \tilde{H}(\BIb,\phib_\dof) = \tilde{H}_0(\BIb)+\eps\,
  \tilde{V}(\BIb,\phib_\dof) \ .
\end{equation} 
The tilde~$\tilde{\ }$ will be dropped henceforth. 
The angles $\phib_1,\ldots,\phib_{\dof-1}$ do not appear in the new
Hamiltonian. Hence, their conjugate momenta $\Ib_1,\ldots,\Ib_{\dof-1}$ are
constants of motion. As functions of the old momenta $\BIa$ only, they are
in involution. The Hamiltonians~(\ref{eq:hama}) and
(\ref{eq:hamb}) are therefore completely integrable. 
Action variables are introduced by means of    
\begin{equation}
\Ic_j = \frac{1}{2\pi}\oint_{\gamma_j}\BIb \, d\Bphib \ , 
\quad j=1,\ldots \dof \ .
\end{equation}
With the transformation~(\ref{eq:trans1}) being unimodular, a set of
fundamental paths $\gamma_j$ on a given invariant torus is determined
by $\phib_i = \const$, $i \neq j$. $\dof-1$ action integrals simply are $\Ic_j
= \Ib_j$ if $j<\dof$, and it remains only one non-trivial integral,
\begin{equation}\label{eq:action2}
        \Ic_N =  \frac{1}{2\pi}\oint\Ib_\dof \, d\phib_\dof \ ,
\end{equation}
where $\Ib_\dof$ is regarded as a function of $\phib_\dof$ and the constants
$\h,\Ib_1,\ldots,\Ib_{\dof-1}$. 

\subsection{EXPANSION AT THE RESONANCE SURFACE}
The resonance surface in $\BIb$-space is given by $\Ib_\dof =
A(\Ib_1,\ldots,\Ib_{\dof-1})$, where $A$ is implicitly defined through
$\dot{\phib}_\dof = \partial H_0 / \partial \Ib_\dof = 0$. 
We expand the Hamiltonian~(\ref{eq:hamb}) on this surface in the direction
perpendicular to it. For this purpose, $|\Ib_\dof-A|$ is assumed to
be of order $\sqrt{\eps}$, as usual in the analysis of resonances; see, e.g.,
\cite{Chir79,LichLieb92}. It is therefore sufficient to consider $V$ as
independent of $\Ib_\dof$ when expanding $H$ in powers of $|\Ib_\dof-A|$ to
quadratic terms 
\begin{eqnarray}
H(\BIb,\phib_\dof) & = & H_0(\Ib_1,\ldots,\Ib_{\dof-1},A) \nonumber\\
\label{eq:Hentw}
& & + \frac12 H''_0(\Ib_1,\ldots,\Ib_{\dof-1},A) (\Ib_\dof-A)^2 \\ 
& & +\eps \, V(\Ib_1,\ldots,\Ib_{\dof-1},A,\phib_\dof) \ , \nonumber
\end{eqnarray}
where the prime $'$ denotes a derivative with respect to $\Ib_\dof$. 
The local approximation~(\ref{eq:Hentw}) is useful provided $H_0'' \neq 0$ is
fulfilled, which calls for a nonlinear dependence of $H$ on $\Ib_\dof$; this
is why it is called ``nonlinear resonance''.    
It is illuminating to have a closer look at $H_0''$ in terms of the old
coordinates 
\begin{equation}\label{eq:h0ss}
 H_0'' = \scp{\m}{\frac{\partial \Bomega_0}{\partial \BIa} \m}\ ,
\end{equation}
where the Jacobian $\partial \Bomega_0/\partial \BIa$ is evaluated on
the resonance surface. Note that since \equ~(\ref{eq:resonanceplane}) holds at
resonance, $\m$ is a tangent to the unperturbed energy surface. Hence, $H_0''$ is
the rate of frequency change on the energy surface perpendicular to the
resonance surface or, equally, a measure of the curvature of the energy
surface in the direction of $\m$. From this it is obvious that systems
with planar unperturbed energy surfaces, like coupled harmonic oscillators,
have to be treated separately.   

Let us consider \equ~(\ref{eq:Hentw}) as a one-degree-of-freedom
Hamiltonian
\begin{equation}\label{eq:H2}
H(\Ib_\dof,\phib_\dof) = \frac{1}{2\M} (\Ib_\dof-A)^2 +
\h_0 + \eps\, V(\phib_\dof) \ .
\end{equation}
The ``mass'' $M = 1/H_0''$ can be positive or negative. The zero level
$\h_0= H_0$ does not affect the dynamics. It is worth mentioning that due to
the ``one-dimensional vector potential'' $A$, the Hamiltonian~(\ref{eq:H2}) is
not invariant under time reversal, which is $\Ib_\dof \rightarrow -\Ib_\dof$
when the other momenta are regarded as fixed parameters.   
We perform a further simplification, usually called ``resonance-centre
approximation", which ignores the dependence of the
potential~(\ref{eq:potential}) upon the constants $\Ib_1,\ldots,\Ib_{\dof-1}$,
i.e. $\tilde{f}(\Ib_1,\ldots,\Ib_{\dof-1},A) \to f$. 
Absorbing $|f|$ in the perturbation parameter $\eps>0$ and 
using the freedom of shifting the angle $\phib_\dof \to \phib_\dof +
\const$, we take  
\begin{equation}\label{eq:potential2}
V(\phib_\dof) = \sign(\M f)\cos{(q\phib_\dof)} \ .
\end{equation} 
With $\Es = (\h-\h_0)\,\sign(\M)$ as the
energy of the $\dof$th degree of freedom, we get from energy
conservation $H(\Ib_\dof,\phib_\dof) = \h$  
\begin{equation}\label{eq:eIb0}
  \Ib_\dof = \pm \sqrt{2|\M|[\Es-\eps \cos{(q\phib_\dof)}]} + A \ .
\end{equation}    
This is the textbook planar pendulum if $q=\eps=1$ and $A = 0$; see, e.g.,
\cite{LichLieb92} and for a quantum mechanical treatment
see~\cite{AldrFerr80}. Inspection of \fig~\ref{fig:portrait}a reveals that for
general parameters the phase portrait differs from that of a pendulum in that
there is a chain of $q$ identical islands centred at $\Ib_\dof = A$
instead of just a single island centred at $\Ib_\dof = 0$. 
Let us specify briefly the invariant curves. At a fixed energy $\Es > \eps$, 
there exist two invariant circles representing rotational-like motion (not
necessarily related to physical rotations) with opposite sense of rotation
$\patch = \sign(\dot{\phib}_\dof) = \pm 1$, where $\dot{\phib}_\dof = \Ib_\dof
- A$. These circles are smooth deformations of the
unperturbed ones. All other invariant objects, usually subsumed under
the term ``isolated resonance zone'', are created by the perturbation:    
a separatrix and $q$ unstable (hyperbolic) equilibrium points at the critical energy $\Es = \eps$; 
$q$ invariant ``island circles'' representing oscillatory motion in one of
the potential wells labelled by $\Kindex = 1,\ldots,q$ at fixed $\Es \in
(-\eps,\eps)$;  
$q$ stable (elliptic) equilibrium points at the lowest
energy $\Es = -\eps$.   
It is to emphasize that the phase-space embedding of the island tori is
topologically different from that of the smoothly deformed tori.

The system exhibits $2q$ discrete symmetries. It is invariant under reflections
with respect to $\phib_\dof = (\Kindex-1)\pi/q$, $\Kindex = 1,\ldots,q$. These
symmetries can be removed by implementing elastic reflections
$\dot{\phib}_\dof \rightarrow -\dot{\phib}_\dof$ about $\phib_\dof = 0$ and
$\phib_\dof = \pi/q$ as depicted in \fig~\ref{fig:portrait}b. This symmetry
reduction restricts the oscillations to one half of the first potential well,
whereas it converts the rotations to oscillations within the interval
$[0,\pi/q]$.   
\def\figportrait{%
a) Sketch of an isolated resonance zone with $q=2$ in the $(\Ib_\dof,\phib_\dof)$-plane; see \equ~(\ref{eq:eIb0}). 
The lines $\phib_\dof = 0$ and $\phib_\dof = 2\pi$ are identified.  
Shaded regions represent action integrals.
Filled circles and squares mark stable and unstable equilibrium points,
respectively. 
The thick dashed line is the integration path $C_\Theta$ of the tunnel
integral specified in \sect~\ref{sec:quantum}.
b) Symmetry reduced resonance. 
Dashed lines symbolize elastic reflections. 
}
\def\FIGportrait{\centerline{\psfig{figure=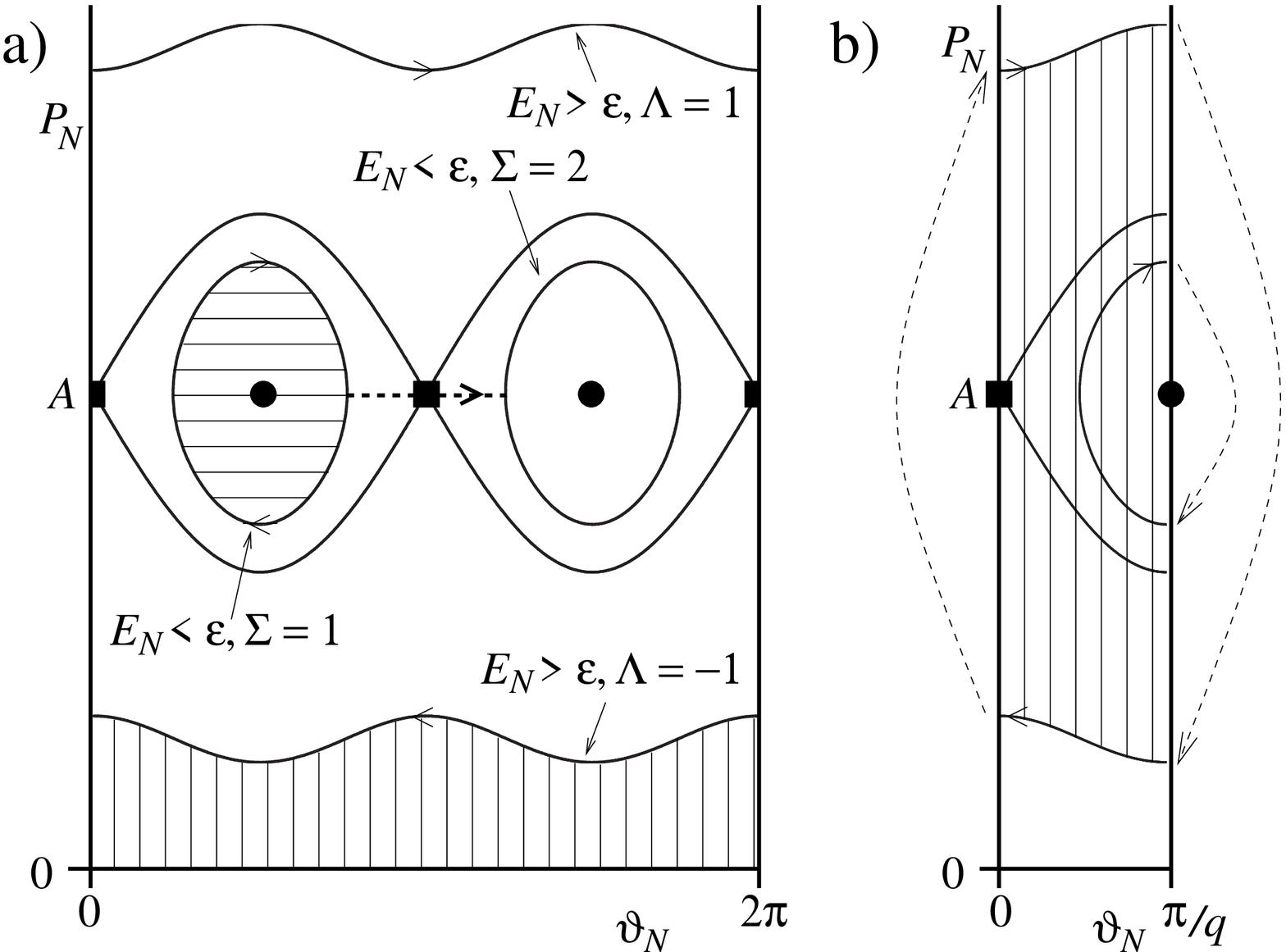,angle=0,width=8.5cm}
}}
\FIGo{fig:portrait}{\figportrait}{\FIGportrait}

\subsection{ACTIONS AND ENERGY SURFACES}
We now turn to the calculation of action variables. 
At fixed $\Es > \eps$ and fixed $\patch$, the action
integral~(\ref{eq:action2}) is the area in the $(\Ib_\dof$,
$\phib_\dof)$-plane between the invariant circle and the line $\Ib_\dof = 0$
as illustrated in \fig~\ref{fig:portrait}a. We choose the closed integration
path to be parametrized by $\phib_\dof$ increasing from $0$ to $2\pi$. The
action $\Ic_\dof$ takes on positive as well as negative values 
depending on the sense of rotation $\patch = \pm 1$ and, remarkably, on $A$.  
At fixed $\Es < \eps$, $\Ic_\dof$ is the area enclosed by the
invariant circle. The chosen integration path lies inside a given potential
well going from the left turning point ($\dot{\phib}_\dof = 0$) to the right
one (along the $\dot{\phib}_\dof > 0$-branch) and back (along the
$\dot{\phib}_\dof < 0$-branch). $\Ic_\dof$ is positive and the same for all
wells. 
Please, pay attention to the fact that $\Ic_\dof$ changes discontinuously upon
traversing the separatrix. As a consequence, there is no unique limiting
action we could assign to the separatrix and the embedded unstable equilibrium
points. Instead, there are three different actions arising from three
different energy limits, namely $\Es \to \eps$ from below and $\Es \to \eps$
with $\patch = \pm 1$ from above.  
This is in strong contrast to the continuous behaviour of the action
$\Icred_\dof$ of the symmetry reduced system; see \fig~\ref{fig:portrait}b.
At fixed $\Es < \eps$, the integration path goes from the left turning point
to the ``solid wall'' at $\phib_\dof = \pi/q$ and back after being
reflected. With increasing energy, the left turning point wanders towards
$\phib_\dof = 0$. Upon crossing the separatrix the smooth turning point is
replaced by a reflection at $\phib_\dof = 0$. This does not spoil the 
continuity of $\Icred_\dof$ (but its smoothness) since at $\Es > \eps$ the
action integral is the area between the two branches of the invariant circle
($\Ib_\dof>A$ and $\Ib_\dof<A$) as illustrated in
\fig~\ref{fig:portrait}b. 
Note that $\Icred_\dof$ does not depend on $A$ as opposed to $\Ic_\dof$. This
is related to the remarkable fact that the symmetry reduction here does not
only reduce phase space area by a factor, it also shifts its value by a
constant. We call this a ``non-trivial symmetry reduction''.   

The calculation of the action integrals is straightforward, and gives 
\renewcommand{\arraystretch}{1.4}
\begin{equation}\label{eq:actions}
  \Icred_\dof(\Es) = \left\{
  \begin{array}{cl}
  \frac{2\sqrt{2}}{q\pi}\sqrt{|\M|}
  \sqrt{\Es+\eps}\;\Eell(1/k) & \mbox{if} \quad \Es > \eps \\
  \sqrt{\eps}\frac{4}{q\pi}\sqrt{|\M|}
  \left[\Eell(k)-(1-k^2)
  \Kell(k)\right] & \mbox{otherwise} 
  \end{array}\right.
\end{equation} 
\stdarray
and  
\begin{equation}\label{eq:actions2}
    \Ic_{\dof}(\Es) = \left\{
  \begin{array}{cl}
\patch q \Icred_\dof + A & \mbox{if} \quad \Es > \eps \\
2 \Icred_\dof & \mbox{otherwise} ,
  \end{array}\right.
\end{equation} 
where $\Kell(k)$ and $\Eell(k)$ are the complete elliptic integrals of first
and second kind in the notation of \cite{GradRyzh65,Byrd71}, 
with modulus $k^2 = (\Es/\eps+1)/{2}$. 
Let us now come back to the $\dof$-degrees-of-freedom system by noting
that $\Icred_j = \Ic_j = \Ib_j$ if $j < \dof$ (we do not care about further
possible discrete symmetries). The system has the one-component property
stemming from the continuity of $\Icred_\dof$. This is mirrored by
the geometrical fact that the two different regions of $\BIcred$-space, the
interior of the resonance zone with $\Es(\BIcred) < \eps$ and the exterior
with $\Es(\BIcred) > \eps$, are continuously connected at the ``separatrix
surface'' $\Es(\BIcred) = \eps$.     
The situation is more involved in $\BIc$-space. The interior of the resonance
zone consists of $q$ identical parts labelled by $\Nindex = 1,\ldots,q$. The
exterior is made of two parts labelled by $\patch = \pm 1$ which are separated
by a gap with size proportional to the square-root of the perturbation
parameter, 
\begin{equation}\label{eq:DIc}
 \Delta\Ic = 2q\Icred_\dof(\Es = \eps) = \sqrt{\eps|\M|}\frac{8}{\pi} \ .
\end{equation}
A special situation occurs in the limit of vanishing perturbation
strength. The resonance zone disappears, whereas its exterior coincides with
the unperturbed action space after applying the inverse of
transformation~(\ref{eq:trans1}). It is therefore reasonable to subject this
region to the inverse transformation also for finite perturbation, while
keeping the other region as it is. This changes the set of fundamental paths
only in the former region, which is allowed because the other region is
separated by a separatrix which prevents a smooth continuation of fundamental
paths anyway. We thus introduce new action variables as
\begin{equation}\label{eq:trans3}
\BId = \left\{
  \begin{array}{ll}
  \BIc            & \mbox{inside the resonance zone} \\
  \mmatrix^t \BIc & \mbox{outside.}  
  \end{array}\right.
\end{equation} 
In $\BId$-space, the gap between the two parts outside the resonance zone is  
\begin{equation}\label{eq:DId}
 \Delta\Id = |\m|\Delta\Ic = \sqrt{\eps|\M|}|\m|\frac{8}{\pi} \ .
\end{equation}
Comparison with \equ~(\ref{eq:h0ss}) brings to light that $\Delta\Id$ does not
depend on the length of $\m$. Roughly speaking, $\Delta\Id$ is small (large)
if the curvature of the energy surface in $\m$-direction is large (small).

The frequencies $\Bomega = \partial H/\partial \BIc$ are calculated
analogously as the actions giving 
\renewcommand{\arraystretch}{1.4}
\begin{equation}
\omega_\dof = \sign(\M)\left(\frac{\partial \Ic_\dof}
{\partial \Es}\right)^{-1}
\end{equation}
and
\begin{equation}
  \omega_j = \left\{
  \begin{array}{ll}
  \frac{\partial \h_0}{\partial \Ic_j} 
  -(\Ic_\dof-A)\frac{\omega_\dof}{2\M}\frac{\partial\M}{\partial \Ic_j} 
  - \omega_\dof \frac{\partial A}{\partial \Ic_j} 
  & \mbox{if} \quad \Es > \eps \\
  \frac{\partial \h_0}{\partial \Ic_j} 
  -\Ic_\dof\frac{\omega_\dof}{2\M}\frac{\partial\M}{\partial \Ic_j}     
  & \mbox{otherwise} 
  \end{array}\right.
\end{equation}
\stdarray
with $j < \dof$.

\subsection{EXAMPLE: COUPLED ROTORS}
Let us illustrate the previous considerations with an example of $\dof$ coupled
identical rotors described by the Hamiltonian
\begin{equation}\label{eq:ehama}
  H(\BIa,\Bphia) = \frac{1}{2}\scpi{\BIa} +\eps \cos{(q\scp\m{\Bphia})} \ .
\end{equation} 
There is no need to expand this function; the unperturbed Hamiltonian
$H_0 = \scpi{\BIa}/2 = \scp{\BIa}{\BIa}/2$ is already a polynomial of second
degree in $\BIa$ and the perturbation is independent of $\BIa$. 
The space of the unperturbed action $\BIa$ is foliated by
$(\dof-1)$-dimensional concentric energy spheres $H_0(\BIa) = \h$. The
resonance surfaces $\scp\m{\Bomega_0} = \scp\m{\BIa} = 0$ form a dense set of
$(\dof-1)$-dimensional planes passing through the origin. 
Figure~\ref{fig:esurfa}a shows such surfaces for two degrees of
freedom. The energy surface provides a comprehensive picture of the
dynamics (free motion of two particles with coordinates $\phia_1$ and
$\phia_2$ on a circle) at a fixed energy, in contrast to phase portraits as in
\fig~\ref{fig:portrait} which contain only information about a single degree
of freedom.  
The one-piece energy surface is present in all four quadrants of action
space. From this we can infer the existence of only a single type of
motion with two rotational degrees of freedom. 
Each point on the energy surface corresponds to an invariant 2-torus (a
two-dimensional torus) in phase space with the outward normal being the
torus' fundamental frequencies. The most important information contained in
the two frequency components is their ratio, the winding number.   
A rational winding number indicates that the periods of the rotors are
rationally related. For example, a $(-1,1)$-resonance implies identical
periods, so the particles move around the circle synchronously. In phase
space, this motion is a periodic orbit. The time-independent
phase difference $\phia_2-\phia_1$ parametrizes a one-parameter family of such
periodic orbits, forming a resonant 2-torus. These resonant tori
are located on resonance surfaces in action space. For two degrees of freedom,
the resonance surfaces are one-dimensional (see \fig~\ref{fig:esurfa}a) but
we nevertheless refer to them as ``surfaces''.   
Resonances are of great importance due to their sensitivity to perturbations. Under
general perturbations, chaotic motion spreads out from resonances (and
separatrices).  
\def\figesurfa{%
(a) Energy surface $\h = 1/2$ and $(-1,1)$-resonance surface of two free
rotors in $\BIa$-space; (b) transformed energy and $(0,1)$-resonance surface
in $\BIb$-space.}
\def\FIGesurfa{\centerline{\psfig{figure=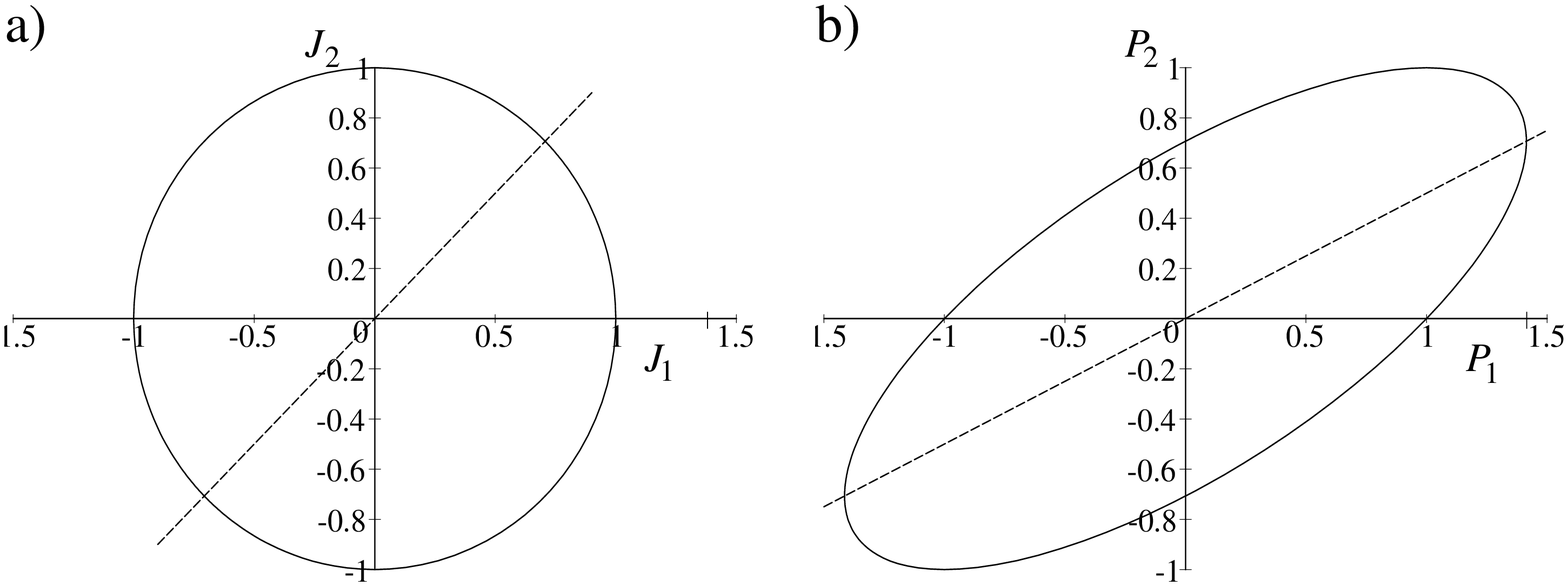,angle=0,width=15.0cm}
}}
\FIGo{fig:esurfa}{\figesurfa}{\FIGesurfa}

For some special $\m$ with $q=1$, the coupled-rotor model has a simple
physical interpretation. First, if all numbers $m_1,\ldots,m_\dof$ are  
zero except $m_j$, the perturbation can be regarded as an harmonic
potential of a spring connecting particle $j$ with a fixed point on the
circle. Note that the spring has a ``negative spring constant'' for positive
$\eps$. Second, $m_i = 1$, $m_j = -1$ and all other components
vanishing models a spring between particle $i$ and particle $j$.   
We discuss this case for two degrees of freedom in more
detail. In order to have a unimodular transformation~(\ref{eq:trans1}) with
$\m = (-1,1)$, we choose $(\dio_1,\dio_2) = (0,1)$ leading to $\M = 1/2$, $A =
\Ib_1/2$, and $\h_0 = \Ib_1^2/4$. The new angle $\phib_1$ is equal to the old
angle $\phia_1$ and $\phib_2 = \phia_2-\phia_1$ describes the relative motion
of the two rotors. It is worth mentioning that the most intuitive transformation,
relative coordinate $\phib_2$ and ``centre of mass'' coordinate $\phib_1 =
\phia_1+\phia_2$ or $\phib_1 = (\phia_1+\phia_2)/2$, is not unimodular. Yet, it
is essential that the transformation~(\ref{eq:trans1}) is unimodular,
otherwise the normals of the transformed energy surface plotted in
\fig~\ref{fig:esurfa}b would not give the fundamental frequencies of the
motion on tori. Note that the $(-1,1)$-resonance ($\dot{\phia}_2-\dot{\phia}_1
= 0$) is transformed into a $(0,1)$-resonance ($\dot{\phib_2} = 0$).  

Both energy surfaces in \fig~\ref{fig:esurfa} are borderless which is quite an
untypical feature in the class of systems studied so far. Ordinary 
energy surfaces of two-degrees-of-freedom systems consist of different patches
bounded by critical points. The critical points are related to isolated
periodic orbits, indicating bifurcations of invariant tori. A critical point
is called elliptic or hyperbolic depending on whether the periodic orbit is 
stable or unstable~\cite{DHJPSWWW97}. At an unstable orbit which is
always accompanied by a separatrix the energy surface has a singular curvature
at the critical point. It is natural that one action is zero at a stable
orbit~\cite{DW94}. This can be achieved by a proper choice
of fundamental paths on the invariant tori.

The perturbed energy surface in $\BIc$-space is calculated according to
\equs~(\ref{eq:actions})-(\ref{eq:actions2}) by fixing the energy $\h$ and
varying the momenta $\Ib_1,\ldots,\Ib_{\dof-1}$. The result shown in
\fig~\ref{fig:esurfb}a is an energy surface which consists of four patches and
is more generic than the unperturbed one.
The isolated resonance zone appears with two small symmetric patches with $|\Ic_1|
\geq \Ic^s = 2\sqrt{\h-\eps}$ which are related by time reversal. A point on
these patches belongs to an island torus where the old angle $\phib_2$
describes oscillations, so only positive values of $\Ic_2$ are meaningful. 
The two hyperbolic points with $(|\Ic_1|,\Ic_2) = (\Ic^s,\Delta\Ic = 
\sqrt{2\eps}\,4/{\pi})$ mark unstable periodic orbits, clockwise and
anti-clockwise rotating with $\phib_2 = \dot{\phib}_2 = 0$,
and separatrix motion, which is asymptotic to the embedded unstable periodic
orbit. The two elliptic points with maximum $|\Ic_1|$ and $\Ic_2=0$
characterize stable periodic orbits, clockwise and anti-clockwise rotating
with $\phib_2 = \pi$, $\dot{\phib}_2 = 0$.   
Outside the resonance zone there exist two patches provided that $\h >
\eps$. If $\h \gg \eps$ as in \fig~\ref{fig:esurfb}a, both patches together
look like the unperturbed energy surface in $\BIb$-space shown in
\fig~\ref{fig:esurfa}b, apart from the gap of size $\Delta\Ic$  where the
unperturbed surface has the 
$(0,1)$-resonance. Points on these patches correspond to rotational motion
similar to the unperturbed motion with both old angles $\phib_1$ and
$\phib_2$ covering the entire interval $[0,2\pi)$. The perturbation just
lifts the constance of the velocities $\dot{\phib}_1$ and $\dot{\phib}_2$.  
The patches are bounded by two pairs of hyperbolic points with 
$\Ic_1 = \pm\Ic^s$. Each pair is related to one of the unstable periodic
orbits and separatrices discussed above.   
Elliptic points do not exist.  
\def\figesurfb{%
Energy surface $\h = 1/2$ of two coupled rotors with $\m = (-1,1)$, $q=1$,
and $\eps = 0.02$ in $\BIc$- (a) and $\BId$-space (b). 
The dashed lines indicate the $(0,1)$- (a) and the
$(-1,1)$-resonance (b) of the unperturbed system. 
The dotted line serves for the construction of the quantity $\Id^k$.}
\def\FIGesurfb{\centerline{\psfig{figure=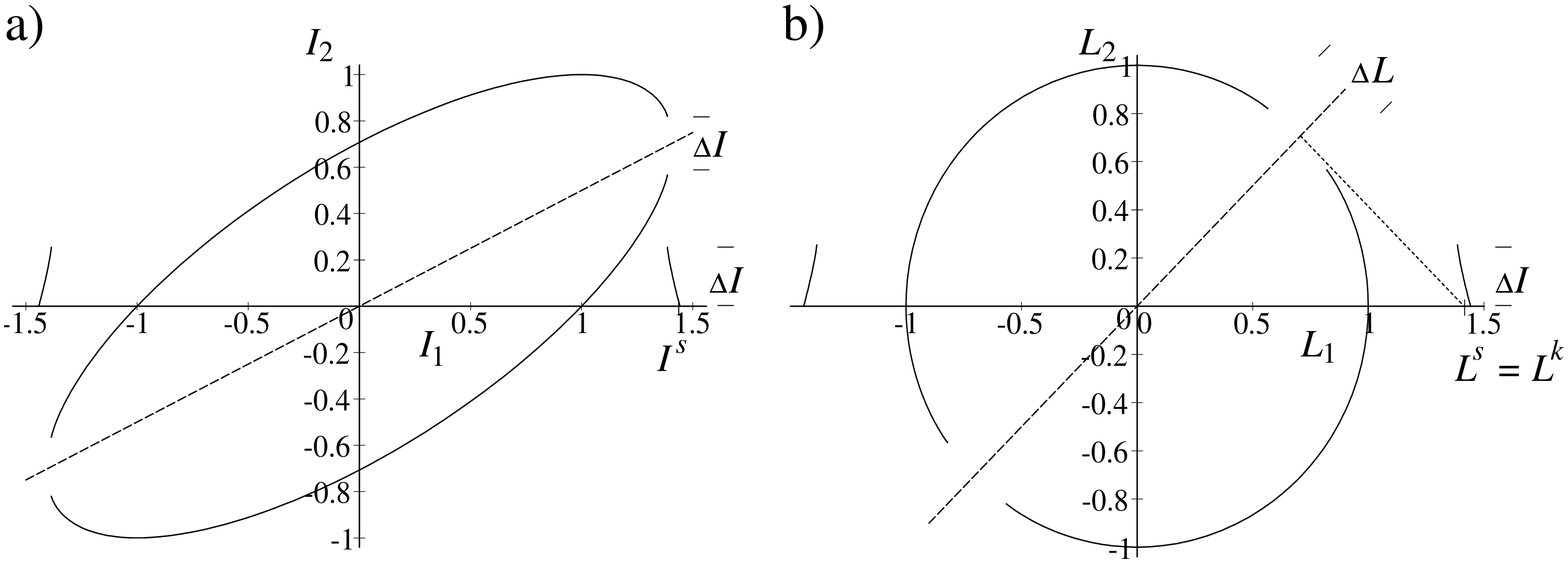,angle=0,width=15.0cm}}
}
\FIGo{fig:esurfb}{\figesurfb}{\FIGesurfb}

Figure~\ref{fig:esurfb}b displays the energy surface after
transformation~(\ref{eq:trans3}) is applied. 
Its rough features are captured by the following slight modification of the
unperturbed energy surface in \fig~\ref{fig:esurfa}a: 
cut in holes of size $\Delta\Id = \sqrt{\eps}\,{8}/{\pi}$ at the
$(-1,1)$-resonance surface; 
draw a line from the intersection point of the resonance and energy surfaces
tangential to the energy surface as pictured in \fig~\ref{fig:esurfb}b; its
intersection point with the $\Id_1$-axis, $\Id^k$, is related to $\Id^s$ via   
\begin{equation}\label{eq:geomkons}
\Id^s = |m_2 \Id^k| \ ;
\end{equation}
add two, almost vertical pieces of height $\Delta\Ic$ at $|\Id_1| = \Id^s =
\Ic^s$.  
In our example with $m_2 = 1$ the quantities $\Id^s$ and $\Id^k$ are
equal. In general, $\Id^s \geq \Id^k$. 

The fine structure of the perturbed energy surface is illustrated in
\fig~\ref{fig:wind} with the help of the winding number.   
The unperturbed winding number $W_0 = \dot{\phia}_2/\dot{\phia}_1$ is simply
$\pm\sqrt{2\h-\Ia_1^2}/\Ia_1$. The $(-1,1)$-resonance is characterized
by $W_0(\h,\Ia_1) = 1$. The same resonance in $\BIb$-space is given by
$W_0(\h,\Ib_1) = \dot{\phib}_2/\dot{\phib}_1 = 0$. 
Comparison of \figs~\ref{fig:wind}b and~\ref{fig:wind}c shows that the
winding number $W(\h,\Ic_1)$ at finite $\eps$ does not differ much from
$W_0(\h,\Ib_1)$, apart from a new piece inside the resonance zone, which takes
on small values of order $\sqrt{\eps}$. 
The behaviour at the separatrix $\Ic_1 = \Ic^s$ can be seen more clearly in
the magnification. Exactly at the separatrix, $W(\h,\Ic_1)$ logarithmically
approaches zero, or, taking a more common point of view, $1/W(\h,\Ic_1)$
diverges logarithmically. This means, on the one hand, that the associated
unstable motion can be regarded as resonant. On the other hand, it means that
even though there is a gap in the energy surface, there is no such gap in the
spectrum of the winding number.    
It can be inferred from \fig~\ref{fig:wind}d that this is also true for the
transformed winding number $W(\h,\Id_1)$. It approaches the value $1$ (in
general $-m_2/m_1$) at the separatrix. 
The derivative of the winding number is large in the vicinity of the
separatrix. As a consequence, there is an accumulation of low-order resonances
($W$ is a fraction of two integers with small denominator) near the separatrix.
\def\figwind{%
Winding number of two coupled rotors with $\h = 1/2$, $\m = (-1,1)$, and
$q=1$. The symmetric regime of negative actions is omitted. 
a) Winding number $W_0(\h,\Ia_1)$ of the unperturbed system. The dotted line
indicate the resonance $W_0(\h,\Ia_1) = 1$. 
b) $W_0(\h,\Ib_1)$.
c) $W(\h,\Ic_1)$ with $\eps = 0.02$. Inset: neighbourhood of the separatrix
$\Ic_1 = \Ic^s$.
d) $W(\h,\Id_1)$.} 
\def\FIGwind{\centerline{\psfig{figure=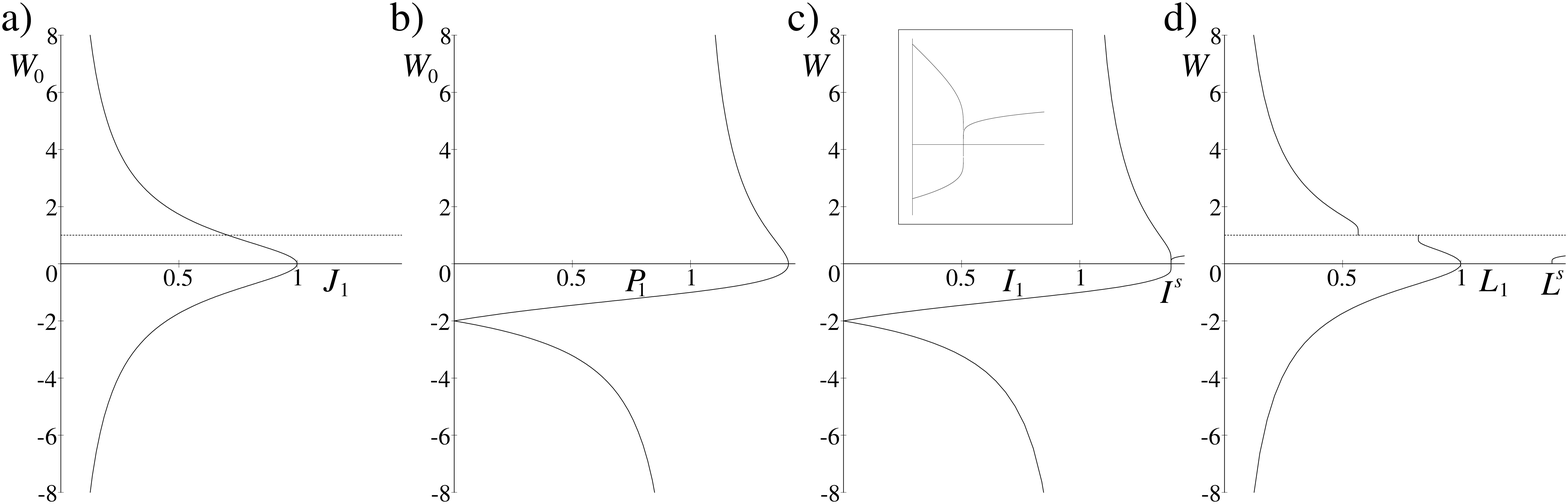,angle=0,width=16.0cm}}}
\FIGo{fig:wind}{\figwind}{\FIGwind}

Figure~\ref{fig:foli} shows how the $\BId$-space is foliated by energy
surfaces. The evident discontinuity differs strongly from the continuity of
$\BIcred$-space, cf. \figs~\ref{fig:foli} and \ref{fig:esurfbred}.  
Let us try to get more familiar with the symmetry reduction on the
basis of the model of two rotors coupled by a spring, $\m = (-1,1)$ and $q=1$. 
The symmetry-reducing reflections about $\phib_2 = 0$ and $\phib_2 = \pi$ are
related to two symmetry transformations, $\phib_2 \to -\phib_2$ and $\phib_2 -
\pi \to \pi-\phib_2$. The first one reads in the old angles:
$\phia_2-\phia_1 \to \phia_1-\phia_2$. This is an interchange of
particle $1$ and $2$, so the reflection at $\phib_2 = 0$ can be
viewed as an elastic reflection between both particles.
The interpretation of the second symmetry transformation,
$\phia_2-(\phia_1+\pi) \to (\phia_1+\pi)-\phia_2$,  is more involved: shift
particle $1$ by $\pi$ on the circle; interchange both particles and finally
shift particle $1$ by $-\pi$. The reflection at $\phib_2 = \pi$ can be seen as
an elastic reflection of particle $2$ with particle $1$ virtually displaced by
$\pi$. 
\def\figfoli{%
Energy surfaces $\h = 1/2, 1/4, 1/8$ of two coupled rotors with $\m =
(-1,1)$, $q=1$, and $\eps = 0.02$ outside (a) and inside (b) the resonance
zone.
The dotted lines mark the separatrix surface.}
\def\FIGfoli{\centerline{\psfig{figure=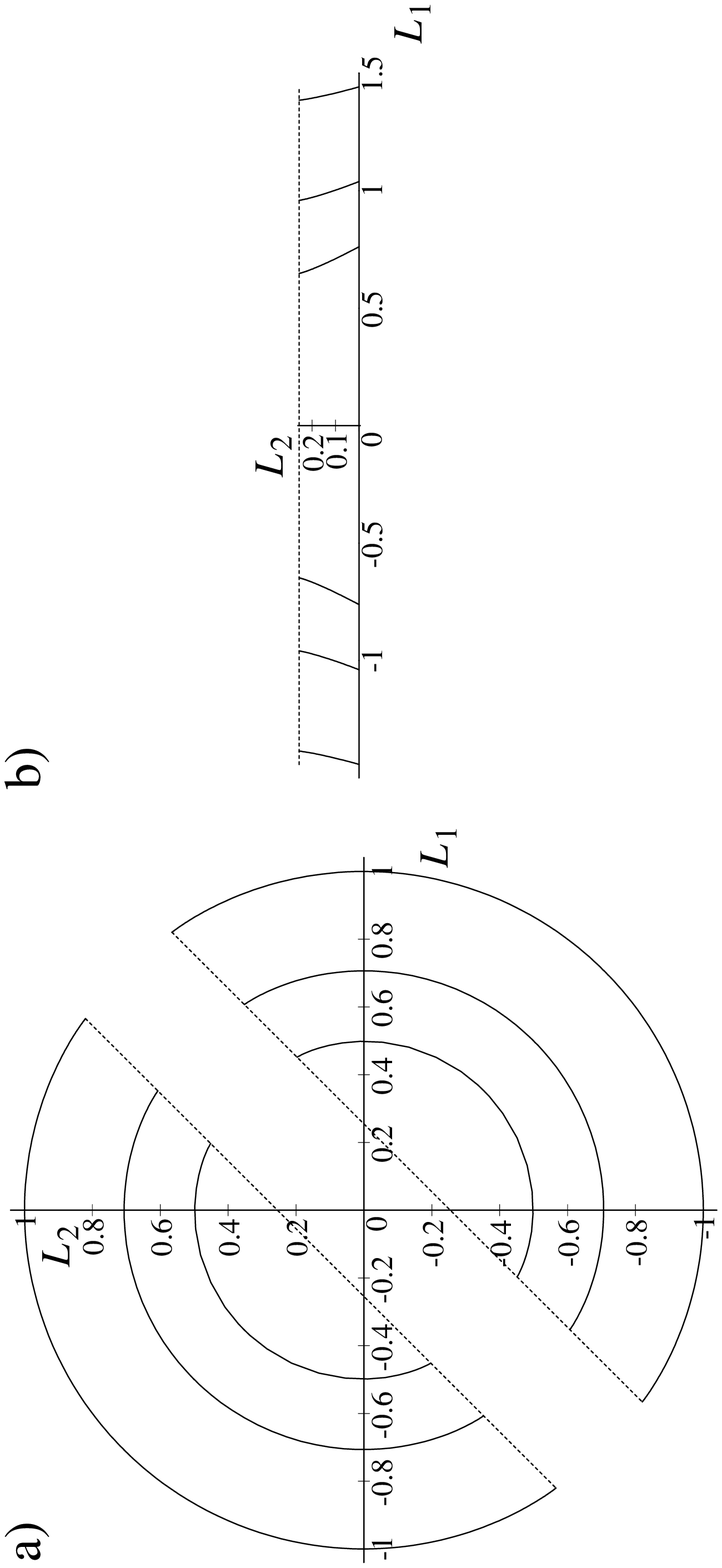,angle=-90,width=15.0cm}}}
\FIGo{fig:foli}{\figfoli}{\FIGfoli}
\def\figesurfbred{%
Energy surfaces $\h = 1/2, 1/4, 1/8$ of two symmetry reduced coupled rotors
with $\m = (-1,1)$, $q=1$, and $\eps = 0.02$. The dotted line marks the
separatrix surface.}
\def\FIGesurfbred{\centerline{\psfig{figure=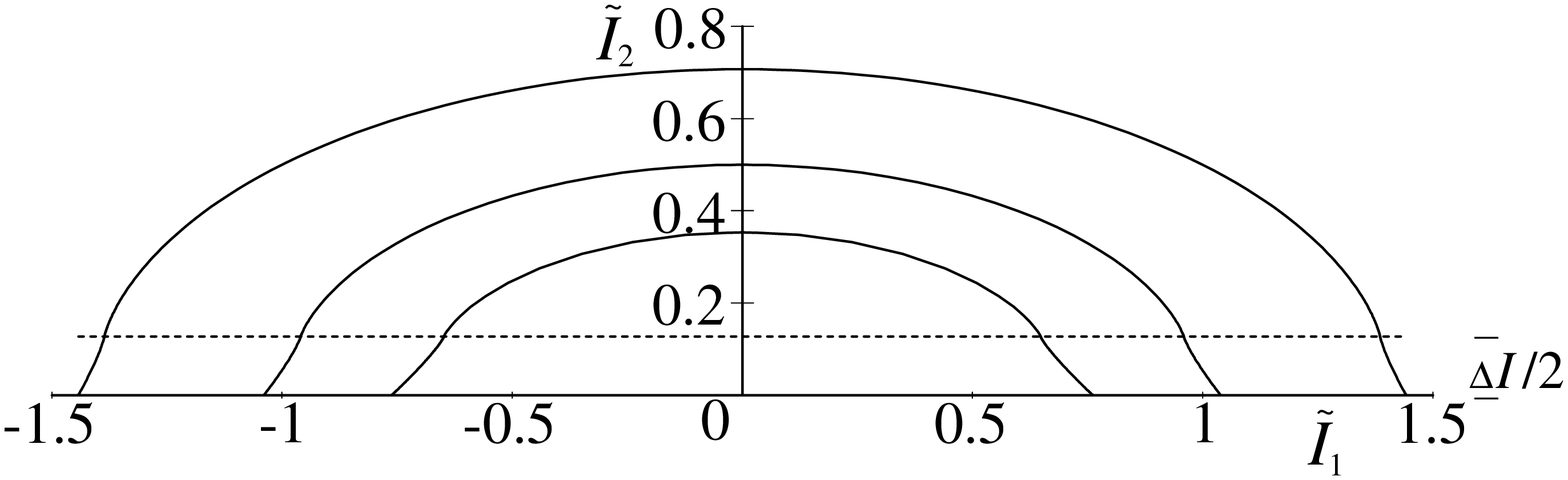,angle=0,width=11.0cm}}}
\FIGo{fig:esurfbred}{\figesurfbred}{\FIGesurfbred}

Let us take a brief look at an example with $m_2,q > 1$, namely $\m = (-2,3)$
and $q=2$ with $(\dio_1,\dio_2) = (1,1)$. We observe from \fig~\ref{fig:d2m23}
that the energy patches are well separated as predicted by
\equ~(\ref{eq:geomkons}). A point within the resonance zone represents two
2-tori, each a combination of a $\phib_1$-rotation and a $\phib_2$-oscillation
in one of the two potential wells.   
\def\figd2m23{%
Energy surface $\h = 1/2$ of two coupled rotors with $\m = (-2,3)$, $q = 2$,
and $\eps = 0.02$. The dashed line marks the $(-2,3)$-resonance of the
unperturbed system.
The dotted line serves for the construction of the quantity $\Id^k$.} 
\def\FIGd2m23{\centerline{\psfig{figure=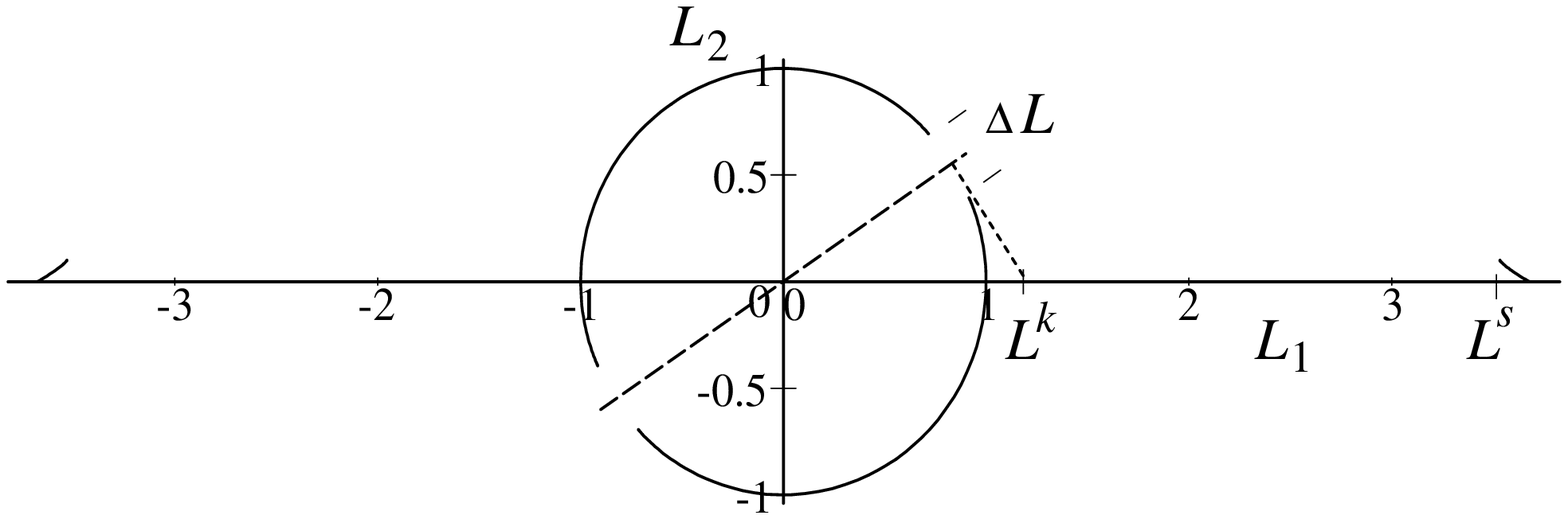,angle=0,width=12.0cm}}}
\FIGo{fig:d2m23}{\figd2m23}{\FIGd2m23}

The description of the two-degrees-of-freedom dynamics extends to three
(and more) degrees of freedom in a natural way. We demonstrate this with three
coupled rotors, using $\m = (1,1,1)$, $q = 1$ and 
$(\dio_1,\dio_2) = (0,1)$.   
Figure~\ref{fig:resa} shows that the
unperturbed energy surface is made of a single piece being present in all
octants of action space, so only one type of motion exists with three
rotational degrees of freedom. Again, the surface is somewhat special in that
it is borderless. Typical energy surfaces of three-degrees-of-freedom
systems are composed of several patches bounded by critical edges. Interior
points of a given patch represent 3-tori. Elliptic edges correspond to stable
isolated 2-tori, whereas hyperbolic edges correspond to unstable isolated
2-tori and separatrices. Corner points represent isolated periodic orbits and
separatrices.   
Although isolated low-dimensional tori do not exist in the free-rotor model,
there are families of low-dimensional tori on resonance
surfaces. Figure~\ref{fig:resa} reveals that these surfaces intersect a
spherical energy surface in great circle meridians. All intersection lines
together form a dense set, the so-called Arnold web. Along 
intersection lines we find resonant 3-tori, one-parameter families of 
2-tori. A torus at an intersection point of two such lines is completely 
resonant, i.e. it is foliated by periodic orbits. The resonance surfaces are
shown in \fig~\ref{fig:resa} up to order one, i.e. $|m_j| \leq 1$. Note that
the order so defined is not invariant under unimodular transformations. The
resonances on the energy sphere in \fig~\ref{fig:resa} and the transformed
surface in \fig~\ref{fig:resb} are therefore not always related via
transformation~(\ref{eq:trans1}). 

Figures~\ref{fig:resc} and \ref{fig:resd} illustrate how the perturbation
modifies the energy surfaces in 3D action space. Let us concentrate on the
$\BId$-representation in the latter figure.
The resonance zone cylindrically surrounds the other parts of the energy
surface. Its elliptic edge in the plane $\Id_3 = 0$ presents isolated
stable 2-tori. Such a torus is essentially a direct product of a circle and
the stable periodic orbit discussed for two degrees of freedom. 
The other edge is hyperbolic and $(0,0,1)$-resonant. Its 
unstable 2-tori and separatrices are again direct products of a circle
with the corresponding two-degrees-of-freedom object.
The resonance zone has no corner points and, correspondingly, no isolated
periodic orbits. 
The energy surface outside the resonance zone is torn open along the
$(1,1,1)$-resonance surface of the unperturbed system;
cf. \figs~\ref{fig:resa} and \ref{fig:resd}. 
Apart from this gap, the shape of the surface is again similar to that of the
unperturbed surface. The hardly visible deformation close to the gap is
uncovered by the graphical representation of resonances.  
The two hyperbolic edges are special $(1,1,1)$-resonances which are
tangentially approached by other resonances. This is analogous to the
earlier-mentioned accumulation of resonances in the vicinity of separatrices
in the case of two degrees of freedom.  
\def\figresa{%
Energy surface $\h = 1/2$ of three free rotors in $\BIa$-space.
Intersection lines with $(m_1,m_2,m_3)$-resonance surfaces, $|m_j| \leq
1$, are shown.}
\def\FIGresa{\centerline{\psfig{figure=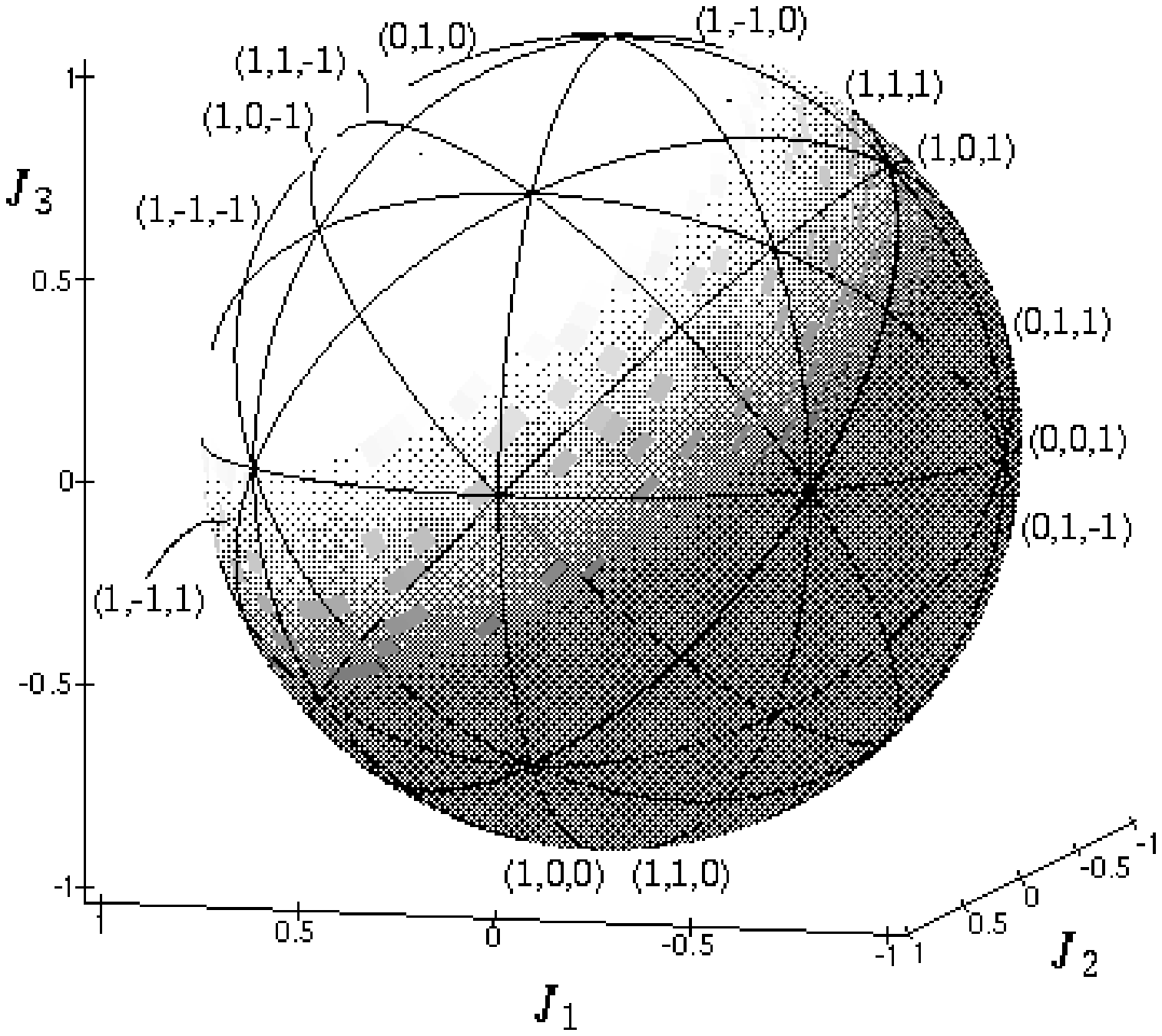,angle=0,width=9.0cm}}}
\FIGo{fig:resa}{\figresa}{\FIGresa}
\def\figresb{%
Energy surface $\h = 1/2$ of three free rotors in $\BIb$-space.} 
\def\FIGresb{\centerline{\psfig{figure=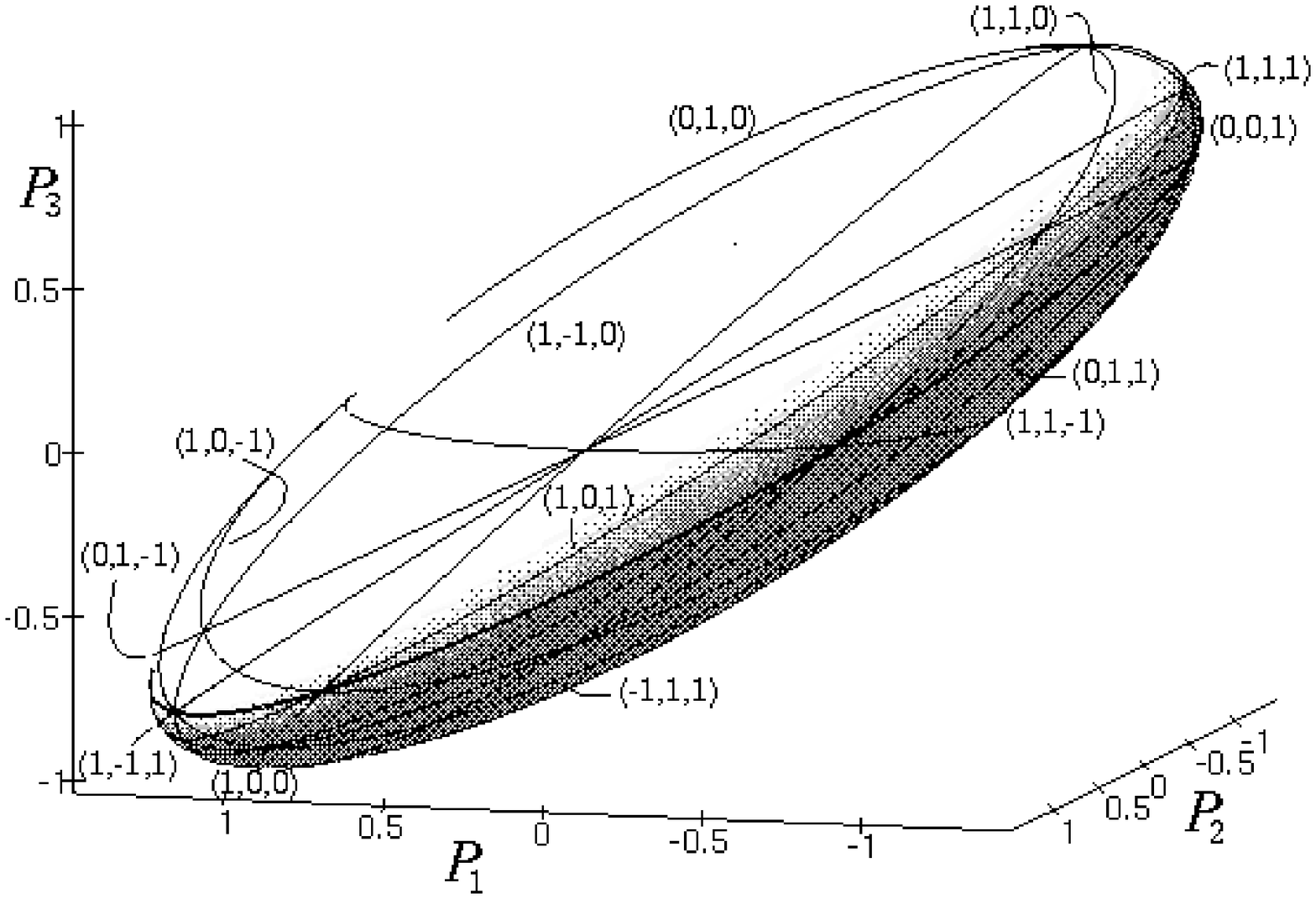,angle=0,width=11.5cm}}}
\FIGo{fig:resb}{\figresb}{\FIGresb}
\def\figresc{%
Energy surface $\h = 1/2$ of three coupled rotors with $\m = (1,1,1)$,
$q=1$, and $\eps = 0.02$ in $\BIc$-space. The unmarked resonances can be
read off from the unperturbed energy surface in \fig~\ref{fig:resb}.}
\def\FIGresc{\centerline{\psfig{figure=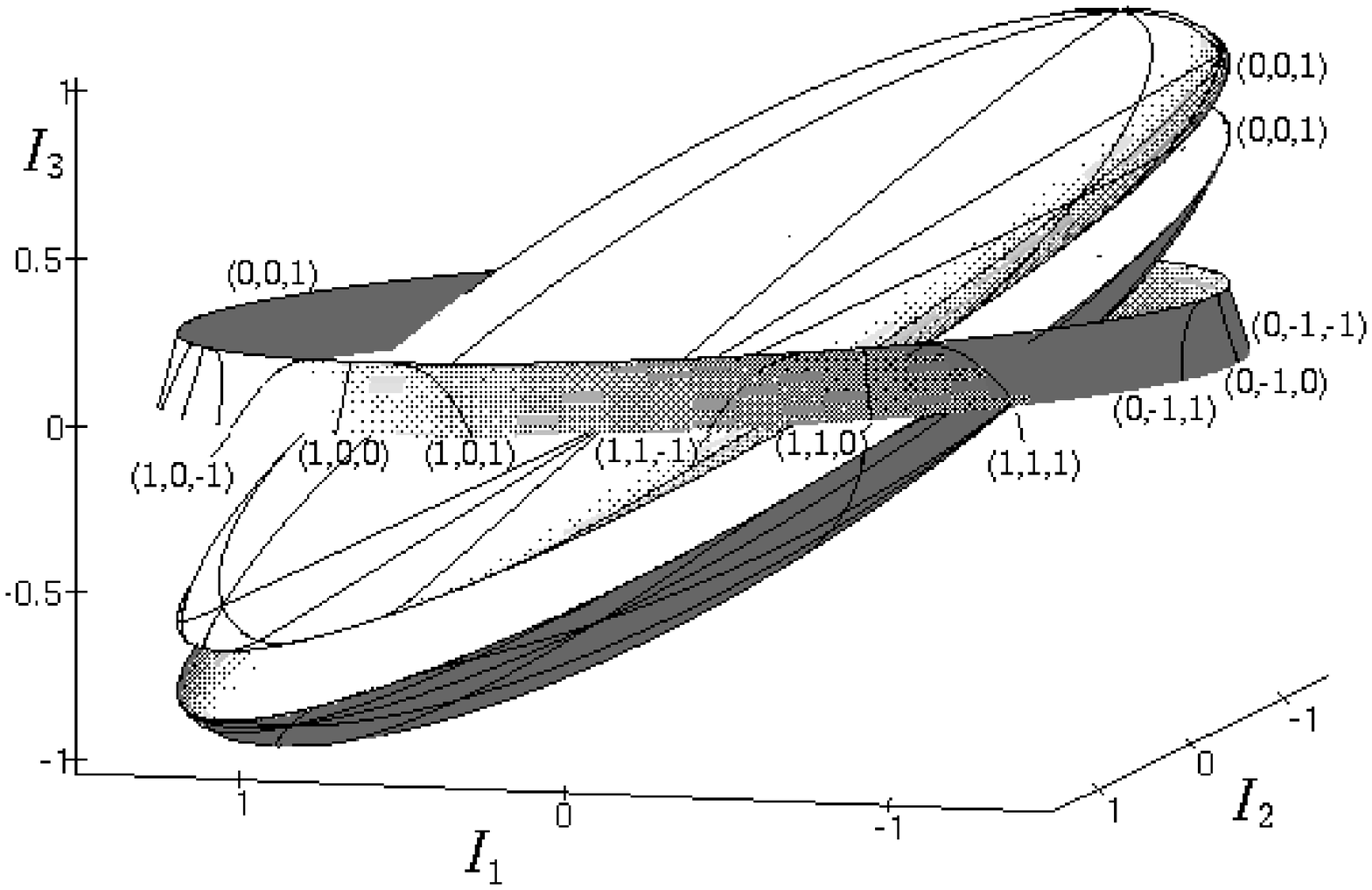,angle=0,width=12.0cm}}}
\FIGo{fig:resc}{\figresc}{\FIGresc}
\def\figresd{%
Energy surface $\h = 1/2$ of three coupled rotors with $\m = (1,1,1)$,
$q=1$, and $\eps = 0.02$ in $\BId$-space. The unmarked resonances can be
read off from the unperturbed energy surfaces in \figs~\ref{fig:resa} and
\ref{fig:resc}.}
\def\FIGresd{\centerline{\psfig{figure=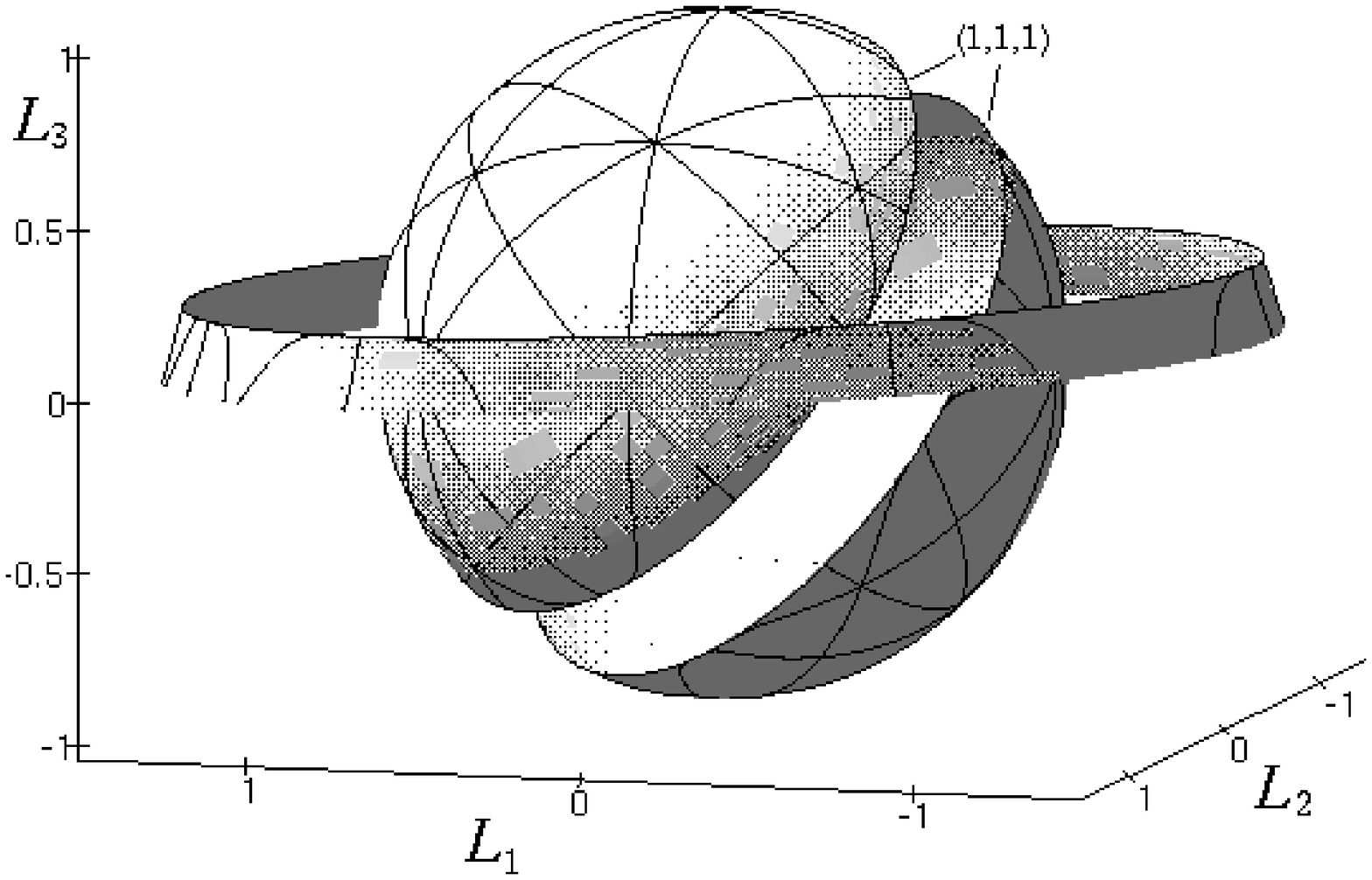,angle=0,width=12.0cm}}}
\FIGo{fig:resd}{\figresd}{\FIGresd}
\clearpage

\subsection{COMPARISON WITH CANONICAL PERTURBATION THEORY}
Having calculated the energy surfaces of coupled rotors exactly, we now
compare them to approximated surfaces. Canonical perturbation theory applies
near-identity transformations parametrized by $\eps$ to the action-angle
variables such that the new Hamiltonian is independent of the angles and
therefore integrable; see, e.g., \cite{LichLieb92}. 
After eighth-order canonical perturbation theory the Hamiltonian~(\ref{eq:ehama}) transforms to
\begin{equation}\label{eq:st}
\Happ(\stIb) =  \frac{1}{2}\scpi{\stIb} 
+ \eps^2\frac14 \xi^2 + \eps^4\frac{5}{64}\xi^6+\eps^6\frac{9}{128}\xi^{10}
+\eps^8\frac{1469}{16384}\xi^{14} \ ,
\end{equation}
with the new actions $\stIb$ and the abbreviation $\xi
=|\m|/(\scp{\m}{\stIb})$. It turns out that $\Happ(\stIb)$ agrees with the
Taylor series of $H(\BId)$ for $\Es > \eps$ which is obtained from expanding
\equs~(\ref{eq:actions})-(\ref{eq:actions2}) in powers of $\eps$,
transforming according to \equ~(\ref{eq:trans3}), and solving (to
eighth order) for the energy.  
Hence, canonical perturbation theory approximates the energy surfaces outside
the resonance zone, cf. \figs~\ref{fig:esurfb}b and \ref{fig:stef}.  
But notice that $\Happ(\stIb)$ diverges on the resonance surface of the
unperturbed system, $\scp{\m}{\stIb} = 0$. This is responsible for the fact
that the approximate energy surfaces do not have hyperbolic boundaries but
instead additional segments close to the resonance surface
without any physical interpretation! 

The energy surfaces within the resonance zone cannot be approximated in this
way. The reason is that the near-identity transformations cannot cope with the
different topology of the phase-space embedding of island tori; see, e.g.,
\cite{LichLieb92}.    
We arrive at the same conclusion by observing that for $\Es < \eps$ the action
$\Ic_\dof$ in \equs~(\ref{eq:actions})-(\ref{eq:actions2}) cannot be expanded
in a Taylor series in powers of $\eps$.     
\def\figstef{%
Energy surface $\h = 1/2$ of two coupled rotors with $\m = (-1,1)$, $q=1$,
and $\eps = 0.02$ in second-order (dotted) and fourth-order (solid)
canonical perturbation theory; compare with the exact surface in
\figs~\ref{fig:esurfb}b and the unperturbed surface in
\fig~\ref{fig:esurfa}a.} 
\def\FIGstef{\centerline{\psfig{figure=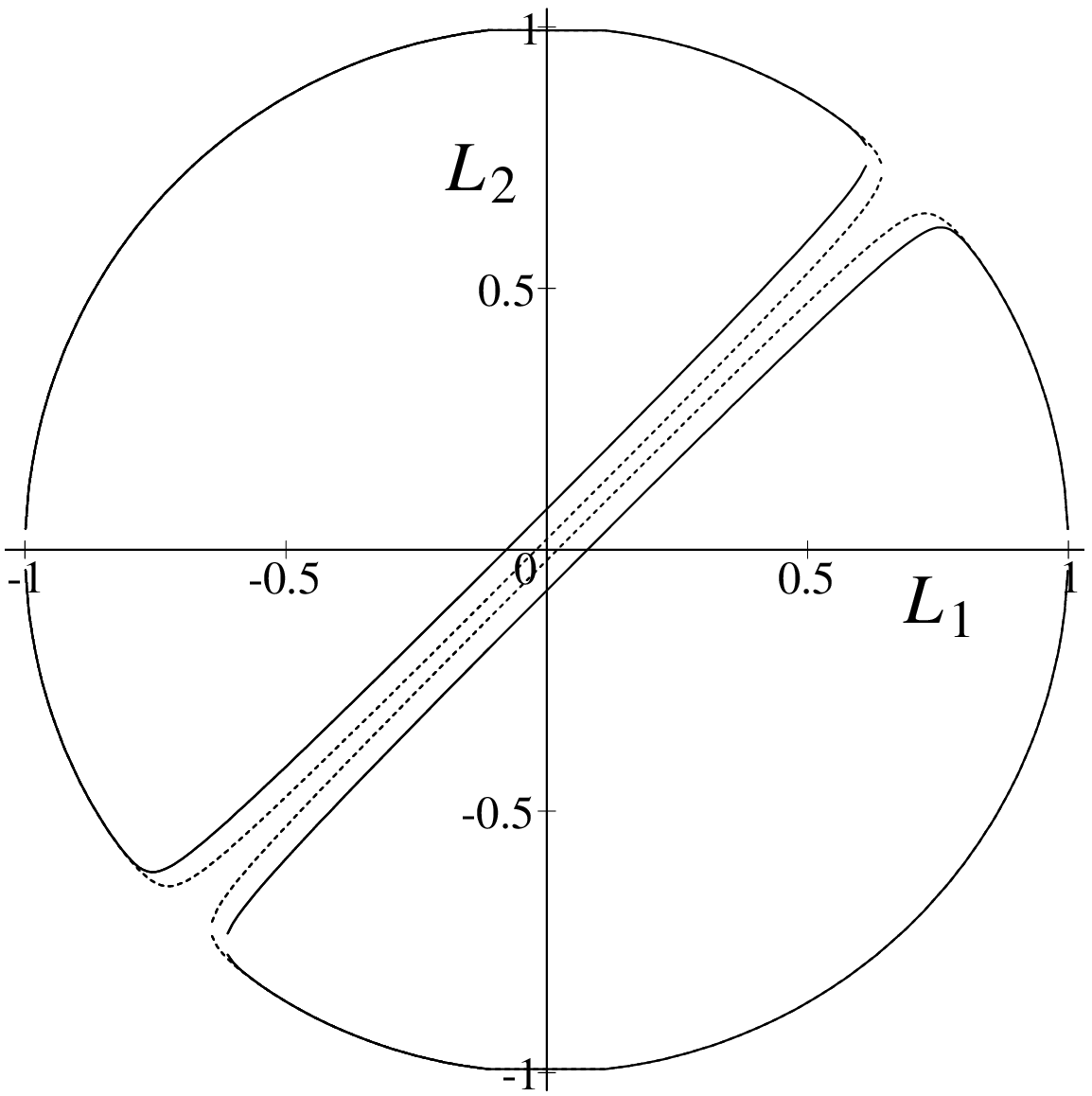,angle=0,width=6.0cm}}}
\FIGo{fig:stef}{\figstef}{\FIGstef}

\section{DISCRETIZATION OF ACTION SPACE}
\label{sec:quantum}
\subsection{SEMICLASSICAL QUANTIZATION CONDITION}
It is instructive to begin with the EBK rule of the unperturbed system,
\begin{equation}\label{eq:ebk2}
\BIa = (\Bna+\Bmaslova/4)\hbar \ ,
\end{equation}
with quantum numbers $\Bna$ and Maslov indices $\Bmaslova$. The latter are
classical indices characterizing the motion on a given invariant
torus: $\maslova_j = 0$ if the $j$th degree of freedom is of rotational type
and $\maslova_j = 2$ if the $j$th degree of freedom is of oscillatory
type~\cite{Keller58,Maslov72}. For rotational motion, $\Ia_j$ and $\na_j$
assume positive as well as negative values (example: angular momentum in a
system with rotational symmetry), whereas for oscillatory motion,
both numbers are usually restricted to non-negative values (example:
one-dimensional harmonic oscillator).   
A Hamilton operator $\hat{H}_0$ with semiclassical eigenfunctions
\begin{equation}\label{eq:wavefunction1}
  \psi_{\na}(\Bphia,\eps=0) \propto \exp[{i\scp{(\Bna+\Bmaslova/4)}{\Bphia}}]
\end{equation}
can be obtained from the unperturbed Hamilton's function $H_0$ with the usual
operator replacement $\Ia_j \rightarrow -i\hbar\partial/\partial\phia_j$,
$j=1,\ldots,\dof$ and $i^2 = -1$. The semiclassical eigenvalues of $\hat{H}_0$
are $\h = H_0(\BIa)$ with $\BIa$ from \equ~(\ref{eq:ebk2}).  
Note that the functions~(\ref{eq:wavefunction1}) are in general not
$2\pi$-periodic in the angles $\phia_j$. This stems from the singularities in
the transformation from the original Cartesian coordinates ${\bm x}$ and
momenta ${\bm p}$ to action-angle variables. 
Transforming the functions~(\ref{eq:wavefunction1}) according to
\equ~(\ref{eq:trans1}) results in 
\begin{equation}\label{eq:wavefunction2}
  \psi_{\nb} (\Bphib,\eps=0) \propto \exp[{i\scp{(\Bnb+\Bmaslovb/4)}{\Bphib}}]
  \ ,
\end{equation}
with new quantum numbers $\Bnb = (\mmatrix^{-1})^t\Bna$ and 
Maslov indices $\Bmaslovb = (\mmatrix^{-1})^t\Bmaslova$. We see here again that
it is important to employ a unimodular transformation; it
guarantees that $\nb_j$ and $\maslovb_j$ are integer-valued and that
consecutive values of $\nb_j$ differ by 1.  

The EBK rule cannot be applied to the perturbed
system~(\ref{eq:hama})-(\ref{eq:potential}) because the actions and Maslov
indices are not globally defined for the entire phase space; their definition
is different for the interior and the exterior of the isolated resonance.
We need a quantization condition which is uniformly valid for the entire
phase space. Our derivation of this uniform quantization
condition is similar to the derivation in~\cite{Ozorio84} but differs in three
respects: an arbitrary number of degrees of freedom is allowed; the
potential~(\ref{eq:potential}) has to be independent of
$\Ib_1,\ldots,\Ib_{\dof-1}$; action-angle variables are used at every stage. 
Let us start from the ansatz  
\begin{equation}\label{eq:wavefunction3}
  \psi_{\nb} (\Bphib) \propto \exp\left[{i\sum_{j=1}^{\dof-1}
  (\nb_j+\maslovb_j/4)\phib_j}\right]\EFpsib{\phib_\dof} \ ,
\end{equation}
with the eigenfunction of the $\dof$th degree of freedom
\begin{equation}\label{eq:wavefunction4}
  \EFpsib{\phib_\dof} = \exp[{i(\maslovb_\dof/4)
    \phib_\dof}]\,\, \EFP{\phib_\dof} \ . 
\end{equation}
The function $\EFP{\phib_\dof}$ is to be $2\pi$-periodic in $\phib_\dof$ in
order to ensure $\EFP{\phib_\dof} \to \exp({i\nb_\dof \phib_\dof})$ and 
$\psi_{\nb} (\Bphib) \to \psi_{\nb} (\Bphib,\eps = 0)$ as $\eps \to 0$. 
Clearly, $\dof-1$ degrees of freedom in this variables are quantized \`a la
EBK 
\begin{equation}\label{eq:qc0}
  \Ic_j= \Ib_j = (\nb_j+\maslovb_j/4)\hbar \, , \quad j=1,\ldots,\dof-1 \ .
\end{equation}
In order to find a quantization condition for the remaining degree of
freedom we insert the conditions~(\ref{eq:qc0}) and $\Ib_\dof \rightarrow
-i\hbar\partial/\partial\phib_\dof$ in Hamilton's function~(\ref{eq:H2})
leading to a Hamilton operator with eigenvalue equation  
\begin{equation}\label{eq:Schroedinger1}
\left[\frac{1}{2\M} \left(-i\hbar\frac{d}{{d\phib_\dof}}-A\right)^2 
+\h_0-\h + \eps V(\phib_\dof)\right]\EFpsib{\phib_\dof} = 0 \ .
\end{equation} 
With the Bloch-wave ansatz
\begin{equation}\label{eq:blochansatz}
\EFpsib{\phib_\dof}=\exp(iA\phib_\dof /\hbar)\, \EFpsic{\phib_\dof}
\end{equation}
a Schr{\"o}dinger equation without ``vector potential'' $A$ is obtained 
\begin{equation}\label{eq:Schroedinger2}
\left[-\frac{\hbar^2}{2\M} \frac{d^2}{d\phib^2_\dof} 
+\h_0-\h +\eps V(\phib_\dof)\right]\EFpsic{\phib_\dof} = 0 \ .
\end{equation} 
The wave functions $\EFpsic{\phib_\dof}$ fulfill ``twisted boundary
conditions'' 
\begin{equation}\label{eq:bc}
  \EFpsic{\phib_\dof+2\pi} = \exp({2\pi i\sigma})\,\, \EFpsic{\phib_\dof} 
\end{equation}
with the real quantity
\begin{equation}\label{eq:sigma}
\sigma = -A(\Ic_1,\ldots,\Ic_{\dof-1})/\hbar+\maslovb_\dof/4 \ ,
\end{equation}
and $d\Psi/d\phib_\dof$ continuous.
With the potential~(\ref{eq:potential2}) ($|f|$ is again absorbed in $\eps$)
we finally get 
\begin{equation}\label{eq:Schroedinger3}
\left[-\frac{\hbar^2}{2|\M|} \frac{d^2}{d\phib^2_\dof} 
-\Es +\eps \cos{(q\phib_\dof)}\right]\EFpsic{\phib_\dof} = 0 \ .
\end{equation}
According to~\cite{Miller68}, the semiclassical solutions $\Es$ of this
eigenvalue problem with boundary conditions~(\ref{eq:bc}) are given by
\begin{equation}\label{eq:qcondition}
  \cos(\phi-\qcf) = 
  \frac{\cos[2\pi (l - \sigma)/q]}{\sqrt{1+\exp({2\Theta/\hbar})}} \ ,
\quad l=1,\ldots,q \ .
\end{equation}
Let us specify the constituents of this formula. The first one is the phase
integral  
\begin{equation}\label{eq:phase}
        \phi(\Es) =  \frac1\hbar\int_{C_\phi}\Ib_\dof \, d\phib_\dof \ .
\end{equation}
At fixed $\Es < \eps$, the integration path $C_\phi$ connects the turning
points inside a potential well. At $\Es > \eps$, the path goes from
$\phib_\dof = 0$ to $\phib_\dof = 2\pi/q$. 
It is an easy exercise to show that the phase integral is related to the
action integrals~(\ref{eq:actions}) by means of $\phi = 2\pi\Icred_\dof/\hbar$.
The path $C_\phi$ lies entirely in the classically allowed region of phase
space, in contrast to the integration path of the tunnel integral 
\begin{equation}\label{eq:tunnelintegral}
        \Theta(\Es) =  -i\int_{C_\Theta}\Ib_\dof \, d\phib_\dof \ ,
\end{equation}
which lies in classically forbidden regions.
At fixed $\Es < \eps$, the path $C_\Theta$ links turning points from
neighbouring potential wells through the barrier as shown in
\fig~\ref{fig:portrait}a. In this case the tunnel integral of the barrier is
positive.  
At $\Es > \eps$, the path connects complex turning points, which are complex
conjugates of each other, yielding a negative tunnel integral. 
Again we show directly the outcome of the calculation:   
\renewcommand{\arraystretch}{1.5}
\begin{equation}
  \Theta = \left\{
  \begin{array}{cc}
  \frac{8\sqrt{\eps}}{q}\sqrt{|\M|}
  k\;\left[\Eell(\sqrt{k^2-1}/k)-\Kell(\sqrt{k^2-1}/k)\right]
  & \mbox{for}\;\Es > \eps \\
  \frac{8\sqrt{\eps}}{q}\sqrt{|\M|}
  \left[\Eell(\sqrt{1-k^2})-k^2\Kell(\sqrt{1-k^2})\right]   
  & \mbox{otherwise,}  
  \end{array}\right.
\end{equation} 
\stdarray
where $k$ is the same modulus as for the action variable~(\ref{eq:actions})-(\ref{eq:actions2}).
$\qcf$ is the ``quantum correction function''
\begin{equation}
\qcf(\Es) =
-\frac{\Theta}{2\pi}\log\left[1+\left(\frac{e\pi\hbar}{4\beta\Theta}\right)^2\right]
\end{equation}
with $e=\exp(1)$ and $\beta = 1.78107$.
A detailed derivation of the formula~(\ref{eq:qcondition}) based on a WKB
ansatz can be found in~\cite{Connor84}. 
Note that even though the WKB method is only valid in the semiclassical limit
$\hbar \to 0$, or to put it another way, for highly excited states, it gives
very often accurate results even for the ground state.

For solving the quantization condition~(\ref{eq:qcondition}) it is convenient
to use a combination of bisection and Newton's method. 
Firstly, the actions $\Icred_1,\ldots,\Icred_{\dof-1}$ are determined from the
given numbers $\nb_1,\ldots,\nb_{\dof-1}$ by virtue of the 
rules~(\ref{eq:qc0}).    
Secondly, the minimum energy $e_0 =
\h_0(\Icred_1,\ldots,\Icred_{\dof-1})-\sign(\M)\eps$ with $\phi(e_0) = 0$ is
computed. 
Thirdly, the running index $j$ is initialized to $0$. 
Fourthly, we look for an energy $e_{j+1}$ such that the interval
$(e_j,e_{j+1})$ includes exactly one eigenvalue. To do so, we note that the
left hand side of condition~(\ref{eq:qcondition}) is essentially a
cosine of $\phi$ ($\qcf$ can be ignored for the following
arguments) and that the right hand side and $\phi$ are
monotonic functions of the energy. Hence, a suitable energy $e_{j+1}$  is given
implicitly by the relation $\phi(e_{j+1}) = j\pi$ which is solved numerically
with Newton's method taking $e_j$ as starting value.  
Having determined the interval, the bisection method is employed to find the
enclosed eigenvalue. 
Finally, $j$ is increased by one and the last steps are repeated until the
desired number of eigenvalues is found.  

\subsection{LATTICE STRUCTURE} 
An analogy between action-space discretization and crystal lattices can be
drawn by rewriting the EBK rule for the unperturbed system~(\ref{eq:ebk2}) as  
\begin{equation}\label{eq:lattice}
\BIa = {\bm a}_0 + \sum_{i=1}^\dof \na_i {\bm a}_i \ ,
\end{equation}
with basis vector ${\bm a}_0 = \Bmaslova\hbar/4$ and primitive lattice
vectors $a_{ij} = \delta_{ij}\hbar$, where $\delta_{ij}$ is the Kronecker
symbol. Equation~(\ref{eq:lattice}) defines a lattice in $\BIa$-space, the
primitive elementary cell of which is an $\dof$-dimensional cube with side
length $\hbar$.    
What kind of lattice does the quantization condition~(\ref{eq:qcondition})
imply? According to~\cite{WWD97}, we formulate the quantization condition in
terms of the action variables of the symmetry reduced system $\BIcred$.
This is trivial for the phase integral $\phi = 2\pi\Icred_\dof/\hbar$, but the
relation $\Es=\Es(\Icred_\dof)$ has to be computed numerically. Note that the
relation is unique and continuous due to the one-component property; see
also~\cite{Wiersig98}. The tunnel integral and the quantum correction 
function become functions of $\Icred_\dof$ via $\Es=\Es(\Icred_\dof)$. We
observe that the quantization condition~(\ref{eq:qcondition}) is a function of
the actions $\BIcred$ alone. 
It is essential to realize that we cannot replace the actions of the symmetry
reduced system $\BIcred$ by the actions of the full system $\BI$ using
\equ~(\ref{eq:actions2}) since the classical index $\patch$ is not provided by
quantum mechanics. Quantum mechanically, we cannot distinguish between a
classical torus with $\patch = +1$ and its symmetric partner with $\patch =
-1$. This is the reason for the importance of the action space of the
associated symmetry reduced system.   

The quantization condition~(\ref{eq:qcondition}) has two solutions with the
same quantum number $l$ in each $\Icred_\dof$-interval of width $\hbar$. We
therefore add two further quantum numbers, $\Pi = \pm 1$ and
$\tilde{\nb}_\dof = 0,1,2,\ldots$. The former distinguishes both solutions and
the latter label the intervals.  
We then verify that the quantization condition can be cast into the form of an
EBK-like rule  
\begin{equation}\label{eq:ebkreddof}
\Icred_\dof = (\tilde{\nb}_\dof+\maslovd_\dof/4)\hbar 
\end{equation}
by inserting this rule into \equ~(\ref{eq:qcondition}). The quantum number
$\tilde{\nb}_\dof$ cancels due to the $2\pi$-periodicity of the
cosine. Inverting the cosine on the correct branch, which is determined by
$\Pi$, gives $2q$ ``Maslov phase functions'' 
\begin{equation}\label{eq:mphasereso}
\maslovd_\dof = \frac2\pi
\mbox{arg}\left(\cos[2\pi (l - \sigma)/q]
+i\Pi\sqrt{\exp(2\Theta/\hbar)+\sin^2[2\pi (l - \sigma)/q]}\right) 
+\frac2\pi\rho \ ,
\end{equation}
where ``arg'' extracts the polar angle $\in [0,2\pi)$ of a complex number.
In contrast to the type of classical motion and its characterizing Maslov index
$\maslov_\dof$ which both change discontinuously at the separatrix, the
Maslov phase function varies smoothly across the separatrix. Hence,
\equs~(\ref{eq:qc0}) and~(\ref{eq:ebkreddof})-(\ref{eq:mphasereso}) do not
define a periodic lattice in the entire action space, instead they define a
``WKB lattice'' in the terminology of~\cite{WWD98}. 

Let us consider first the WKB lattice deep within the resonance zone, $\Es \ll
\eps$. From $\Theta \gg \hbar$ and $\qcf \approx 0$ follows an EBK rule with
constant $\maslovd_\dof = 2-\Pi$. The $l$-independence of $\maslovd_\dof$
manifests itself in a $q$-fold quasi-degeneracy of the eigenvalues
$\Icred_\dof$; $l = 1,\ldots,q$ labels those eigenvalues.  
Translated into action-space geometry, the EBK rules for
$\Icred_1,\ldots,\Icred_\dof$ give two sets ($\Pi = 
\pm 1$) of $q$ identical lattices, each set having unique basis and lattice
vectors. It is possible to combine both sets to a single lattice with a
non-primitive elementary cell, a so-called ``quantum cell'' invented
in~\cite{WWD98}. Here, the quantum cell is a $\dof$-dimensional cube
containing $2q$ quantum states. 
Going back to the full system, one finds from \equs~(\ref{eq:actions2}) and
(\ref{eq:ebkreddof}) 
\begin{equation}\label{eq:ebkresoHO}
\Ic_{\dof} = 2\Icred_\dof = (\nb_\dof+1/2)\hbar 
\end{equation} 
with $\nb_\dof = 2\tilde{\nb}_\dof$ if $\Pi = +1$ and 
$\nb_\dof = 2\tilde{\nb}_\dof+1$ if $\Pi = -1$. We see therefrom that $\Pi$ is
the parity with respect to the potential's symmetry line $\phib_\dof = 
\pi/q$. 
The Maslov index $2$ in the EBK rule~(\ref{eq:ebkresoHO}) is in agreement with
the oscillatory character of the motion of the $\dof$th degree of freedom  in
this phase space regime. 

The other extreme case, $\Es \gg \eps$, well outside the resonance zone,
coincides with the unperturbed limit; we have $\Theta \ll -\hbar$ and
$\qcf \approx 0$ giving 
\begin{equation}\label{eq:maslovdnull}
\maslovd_\dof = \bar{\Pi} \frac4q(l-\sigma) \quad \mbox{with}\quad
\bar{\Pi} = \frac{\Pi}{\sign\{\sin[2\pi(l-\sigma)/q]\}} \ . 
\end{equation}
Eigenvalues $\Icred_\dof$ with different $\bar{\Pi}$ are twofold
degenerated in the non-generic situation of $2\sigma$ being an integer. 
The Maslov phase $\maslovd_\dof$ depends on the actions
$\Icred_1,\ldots,\Icred_{\dof-1}$ via $\sigma(A)$ reflecting the
non-trivial symmetry reduction. Clearly, the quantization
conditions~(\ref{eq:qc0}) and (\ref{eq:ebkreddof}) with
\equ~(\ref{eq:maslovdnull}) give, in general, a non-periodic eigenvalue
distribution in the action space of the symmetry reduced system. Nevertheless,
we will see in the following that periodic lattices in classically small
regions of action space still exist. Each such region, even though classically
small, contains many eigenvalues in the semiclassical limit. Here, $\sigma$ is
approximately a linear function of the actions  
\begin{equation}\label{eq:sigmalin}
\sigma = b_0+\sum_{i=1}^{\dof-1}b_i\Icred_i/\hbar \ .
\end{equation}
Reformulating the quantization conditions~(\ref{eq:qc0}) and
(\ref{eq:ebkreddof}) with the help of \equ~(\ref{eq:sigmalin}) gives an
equation which compares to \equ~(\ref{eq:lattice}) (with $\BIcred$ and ${\bm
{\tilde{\nb}}}$ instead of $\BIa$ and $\Bna$) describing $2q$ lattices with
basis vectors $a_{0i} = \maslovd_i\hbar/4$ and lattice vectors $a_{ij} =
\delta_{ij}\hbar-\delta_{j\dof}\bar{\Pi}b_i\hbar/q$ if $i<\dof$, else $a_{0N}
= \bar{\Pi}(l-b_0)\hbar/q$ and $a_{Nj} = \delta_{Nj}\hbar$. For fixed $\bar{\Pi}$ and
$l$, the lattice vectors have, in general, different lengths and are not
orthogonal; the lattice is $\dof$-dimensional triclinic.   
Note that, in general, a single quantum cell cannot be defined due to
non-matching lattice vectors.   
The reader should realize that the unperturbed eigenvalues of both the full
and the symmetry reduced system lie on a simple $\dof$-dimensional cubic
lattice in the respective ``correct'' action space:  
for the symmetry reduced system, the quantum analog of the classical elastic
reflections are Dirichlet boundary conditions, i.e. vanishing
wave function, on the lines $\phib_\dof = 0$ and $\phib_\dof = \pi/q$
leading to an EBK rule with Maslov index 4 (a hard wall instead of a smooth
turning point increases the Maslov index by 1). 
For the full system, we get an EBK rule from \equs~(\ref{eq:ebkreddof})
and~(\ref{eq:maslovdnull})  using \equ~(\ref{eq:sigma})
\begin{equation}\label{eq:ebkresoorg}
\Ic_{\dof} = \patch q \Icred_\dof + A = (\nb_\dof+\maslovb_\dof/4)\hbar 
\end{equation} 
with the identifications $\bar{\Pi} = -\patch$ and
$\nb_\dof = -\patch q\tilde{\nb}_\dof - l$. In the unperturbed
case, the quantum index $\bar{\Pi}$ therefore equals the classical index $-\patch$
which characterizes the parts outside the resonance zone. It is thus possible
to relate each quantum wave function $\Psi$ uniquely to the classical
$\patch = +1$-region or the $\patch = -1$-region.  
At finite $\Es > \eps$, there is no unique relation between quantum states and
these classical regions as already discussed. When representing the eigenvalues
of the exterior of the resonance zone in the action space of the full system
($\BIc$- or $\BId$-space), we project into the $\patch = +1$-region (we could
also choose the $\patch = -1$-region). In this ``reduced action space'' the
eigenvalues lie on $\dof$-dimensional cubic lattices; see \equs~(\ref{eq:qc0})
and~(\ref{eq:ebkresoorg}).    

\subsection{EXAMPLE: COUPLED ROTORS}
We return to the example of two coupled rotors~(\ref{eq:ehama}) with $\m =
(-1,1)$. The Maslov indices of free rotors, $\maslova_j$, vanish, such 
as the transformed ones, $\maslovb_j$, do. Combining with $A = \Ib_1/2$ we get
$\sigma=-\Ic_1/(2\hbar)=-\nb_1/2$. 
This linear dependence makes for globally periodic lattices outside the
resonance zone in $\BIcred$-space. These lattices with basis vectors
${\bm a}_0 = (0,\bar{\Pi}l\hbar/q)$ and lattice vectors ${\bm a}_1 =
(\hbar,\bar{\Pi}\hbar/(2q))$, ${\bm a}_2 = (0,\hbar)$ are skewed. Because of
the integer-valuedness of $2\sigma$, the lattices with $\bar{\Pi} = +1$ and the
ones with $\bar{\Pi} = -1$ are congruent, or in other words, the eigenvalues
are twofold degenerated.   

Before studying the model with $q = 1$ in detail, it is interesting to mention
that this model has already been treated in Born's 1925 book {\it Vorlesungen
{\"u}ber Atommechanik}~\cite{Born25} with the old Bohr-Sommerfeld quantization
rules (EBK without Maslov indices). This is a too crude approximation as it
will become apparent in the following.   
Consider the right hand side of the quantization
condition~(\ref{eq:qcondition}). It is positive or negative, depending on
whether $\nb_1$ is even or odd. Figure~\ref{fig:bcqa} illustrates the former
situation (for odd $\nb_1$ the non-periodic curve is reflected at the zero
line). The action $\Icred_2$ is discretized according to
\equ~(\ref{eq:ebkresoHO}) in the small-$\Icred_2$ domain (deep within the
resonance zone) and according to \equs~(\ref{eq:ebkreddof}) and
(\ref{eq:maslovdnull}) with twofold degeneracy in the large-$\Icred_2$ domain
(well outside the resonance zone). The transition between these different
kinds of discretization happens in a narrow region around the separatrix with
width of order $\hbar$.     
\def\figbcqa{%
Graphical solution of the quantization condition~(\ref{eq:qcondition}) for two
coupled rotors with $\m = (-1,1)$, $q = 1$, $\eps=0.02$, and $\nb_1$ even.  
Each side of \equ~(\ref{eq:qcondition}) is drawn as a function of $\Icred_2 =
\phi\hbar/(2\pi)$ in units of $\hbar = 0.02$; intersection points indicate  
quantized values of $\Icred_2$. The dashed line marks the separatrix $\Icred_2
= \Delta\Ic/2$.}
\def\FIGbcqa{\centerline{\psfig{figure=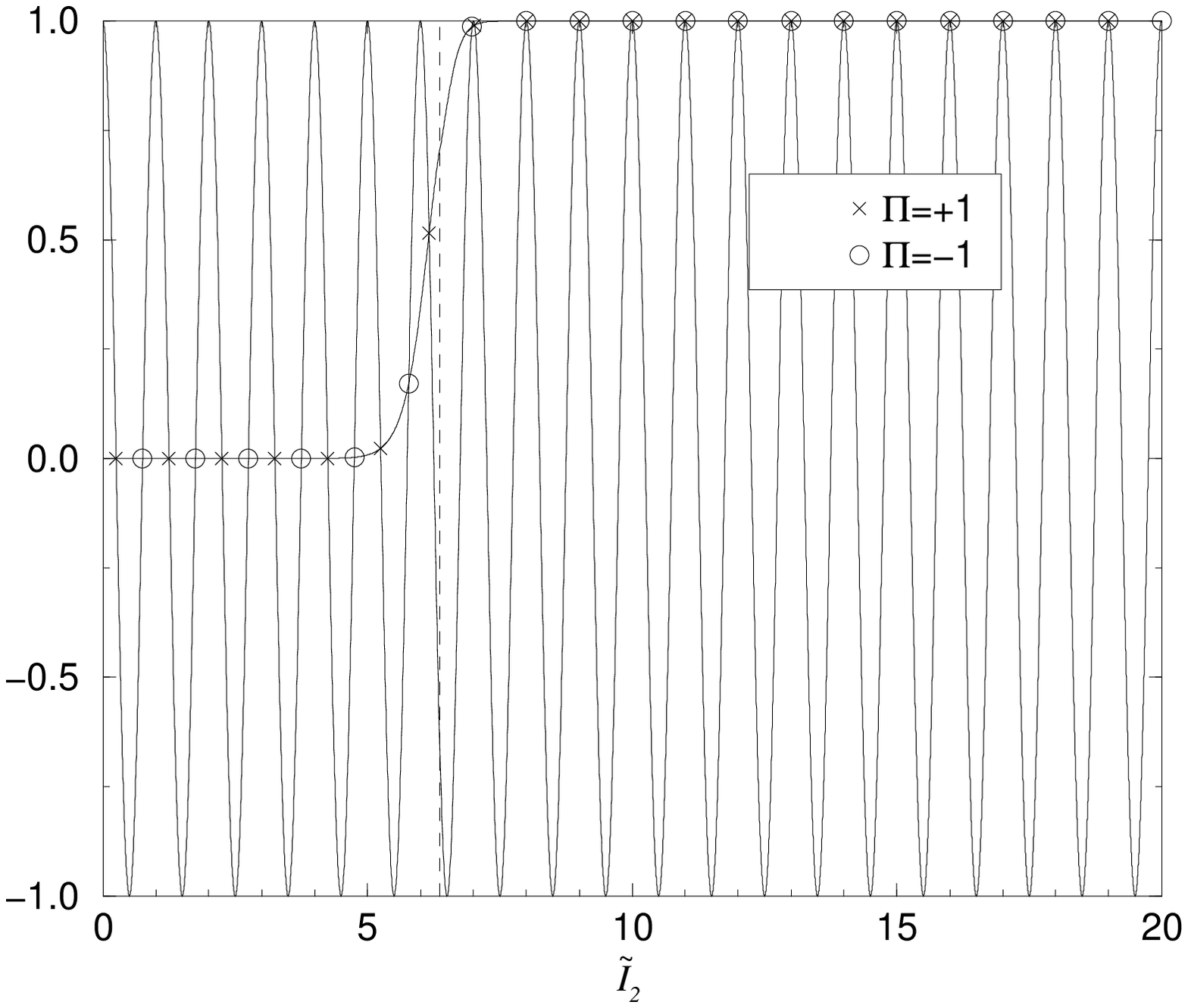,width=10.0cm}}}
\FIGo{fig:bcqa}{\figbcqa}{\FIGbcqa}
 
Solving quantization condition~(\ref{eq:qcondition}) for all values of $\nb_1$
gives the complete set of eigenvalues. Figure~\ref{fig:Xqa} shows their
arrangement in the space of the constants of motion $\h$ and
$\Ic_1=\Ia_1+\Ia_2$. The twofold quasi-degeneracy at large energies is related
to the exact degeneracy of the unperturbed eigenvalues $\h =
(\na_1^2+\na_2^2)\hbar^2/2$ and $\Ic_1 = (\na_1+\na_2)\hbar$. 
The eigenvalue pattern looks rather regular due to the fact that one constant
of motion is an action variable.
However, the underlying regular structures are more transparent in the
$\BIcred$-representation in \fig~\ref{fig:qared}. The eigenvalues are
located on the WKB lattice, which reduces to periodic lattices far away from
the separatrix surface. 
Below the separatrix, i.e. inside the resonance zone, the eigenvalues lie on
two quadratic lattices. It is here trivial to see how a larger square
elementary cell (the quantum cell) could combine both lattices to a single one.
Above the separatrix, i.e. outside the resonance zone, twofold degenerated
eigenvalues lie on two different but congruent skewed lattices, the elementary
cells of which are illustrated in \fig~\ref{fig:qared}. The
integer-valuedness of $2\sigma$ ensures here the existence of a quantum cell,
a body-centred square in the terminology of crystallography.  
As the separatrix surface is crossed from above, a smooth, degeneracy-lifting
transition to the quadratic lattices takes place.  
The lattice outside the resonance zone is even simpler when
projected into the $\patch = +1$-region of $\BId$-space as displayed in
\fig~\ref{fig:qa}. Away from the separatrix surface, 
twofold-degenerate eigenvalues are arranged on quadratic lattices according
to the EBK rule of the unperturbed system~(\ref{eq:qc0}) and
(\ref{eq:ebkresoorg}). The price to pay for recovering a simple lattice for a
subset of eigenvalues is the loss of the coherent picture
of the action-space discretization as shown in \fig~\ref{fig:qared}.
\def\figXqa{%
Semiclassical eigenvalues of two coupled rotors with $\m = (-1,1)$,
$q=1$ and $\eps=0.02$ in the space of the energy $\h$ and action
$\Ic_1=\Ia_1+\Ia_2$ (in units of $\hbar = 0.02$). The symmetric region of
negative~$\Ic_1$ is omitted. }
\def\FIGXqa{\centerline{\psfig{figure=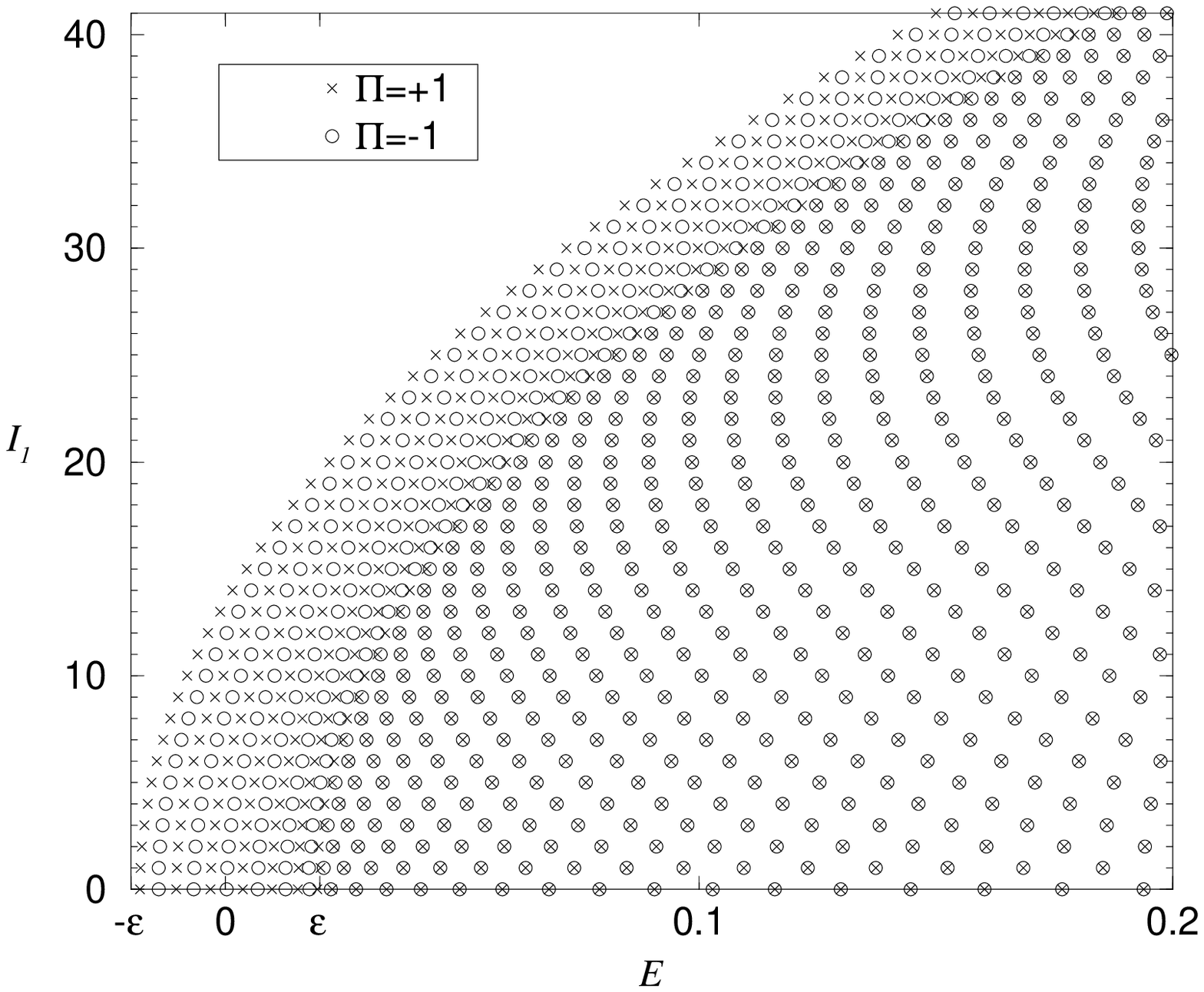,width=11.0cm}}}
\FIGo{fig:Xqa}{\figXqa}{\FIGXqa}
\def\figqared{%
Semiclassical eigenvalues of two coupled rotors with $\m = (-1,1)$,
$q=1$ and $\eps=0.02$ in $\BIcred$-space in units of $\hbar = 0.02$. The
symmetric region of negative action~$\Icred_1$ is omitted.
The solid line is the energy surface $\h=0.15$ and the dashed line is the
separatrix surface.
Filled regions represent elementary cells belonging to $\bar{\Pi} = +1$ (left) and
$\bar{\Pi} = -1$ (right).}  
\def\FIGqared{\centerline{\psfig{figure=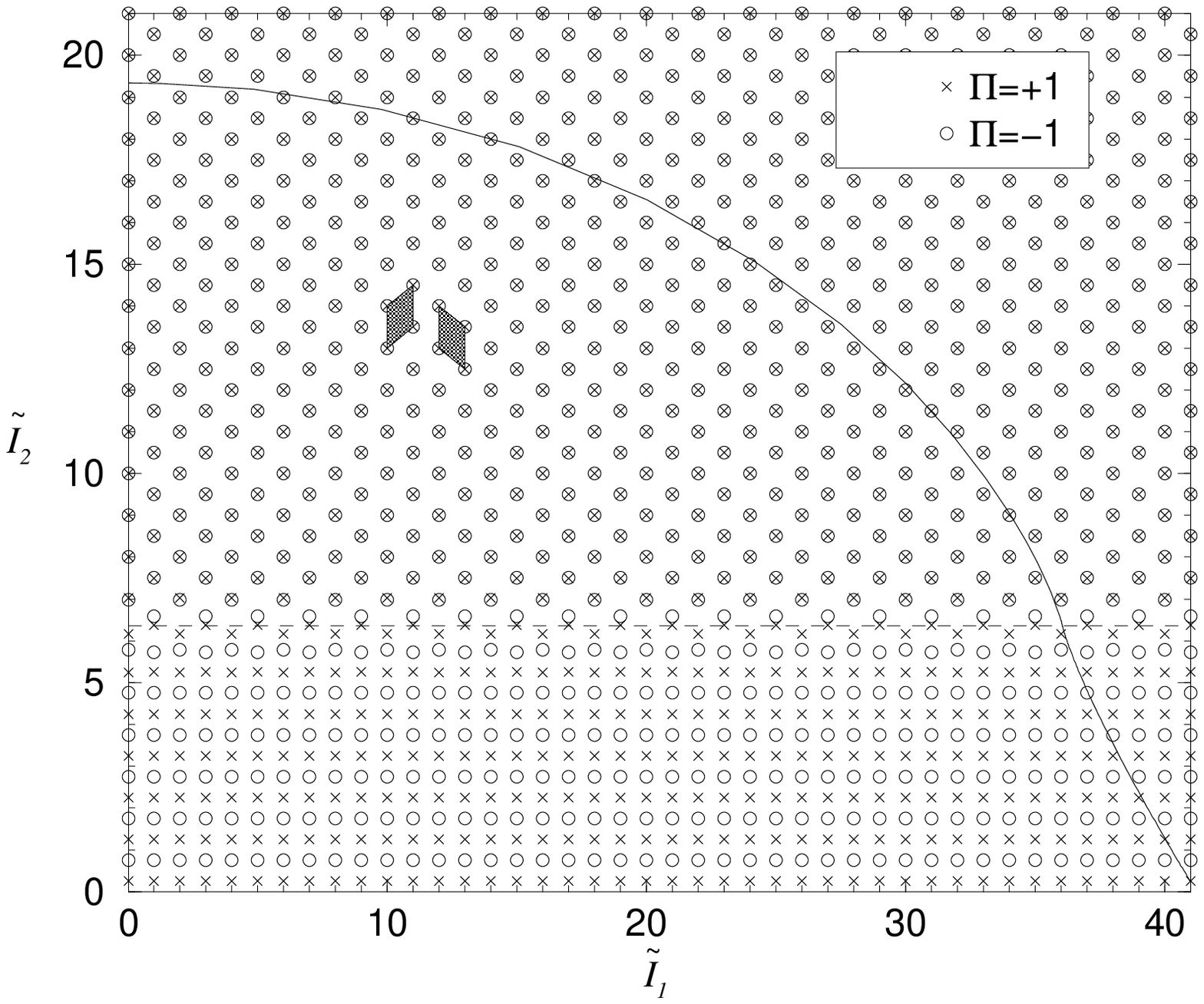,width=11.0cm}
}}
\FIGo{fig:qared}{\figqared}{\FIGqared}
\def\figqa{%
Semiclassical eigenvalues of two coupled rotors with $\m = (-1,1)$,
$q=1$ and $\eps=0.02$ in reduced $\BId$-space in units of $\hbar = 0.02$.
Eigenvalues belonging to the resonance zone are not shown. The solid line
is the energy surface $\h = 0.15$ and the dashed line is the separatrix
surface.}
\def\FIGqa{\centerline{\psfig{figure=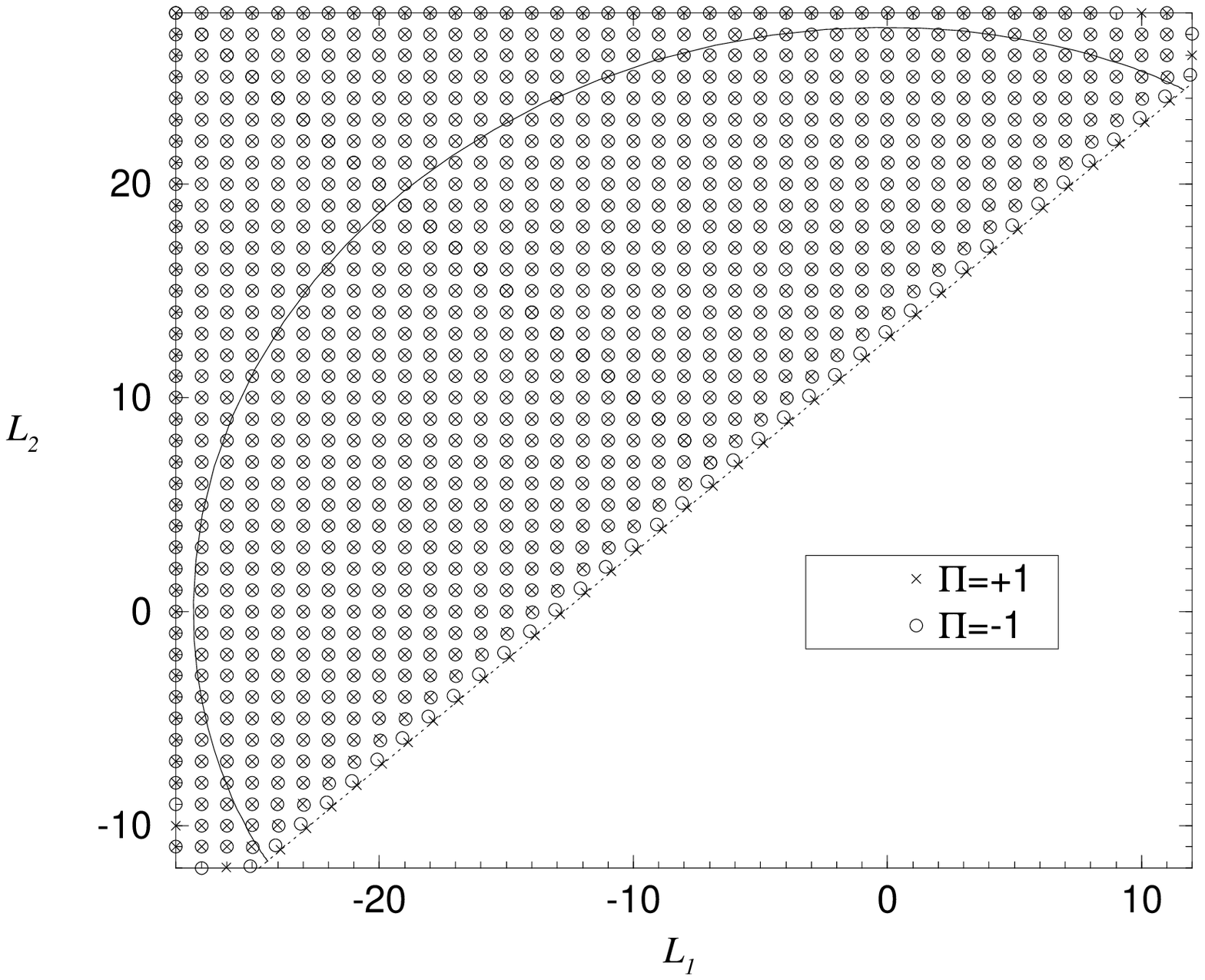,width=11.0cm}}}
\FIGo{fig:qa}{\figqa}{\FIGqa}

Let us now discuss the case $q=2$. 
The quantization condition~(\ref{eq:qcondition}) simplifies to an EBK rule if 
its right hand side vanishes. This is the case if $\nb_1$ is odd since $\sigma
= -\nb_1/2$. In the more complicated case of $\nb_1$ even, we have to
distinguish between $\nb_1/2$ even and $\nb_1/2$ odd. Figure~\ref{fig:bcqb}
illustrates the former situation. Note that the sign of the right hand side of
the quantization condition is determined by $l$. (For odd $\nb_1/2$ the
solutions to $l=1$ and $l=2$ are interchanged.) We see here in addition to
the twofold degeneracy for large $\Icred_2$ also a twofold degeneracy for
small $\Icred_2$.  
Figure~\ref{fig:qbred} shows that above the separatrix surface the
lattice in action space looks similar to $q=1$ in
\fig~\ref{fig:qared}. However, the situation now is actually a bit more
involved: two kinds of elementary cells labelled by $\bar{\Pi} = \pm 1$ are
shifted by $\hbar/2$ in $\Icred_2$-direction ($l=1,2$); we have four lattices
instead of two. Again, a quantum cell could be defined.  
A more striking difference between the case $q=1$ in
\fig~\ref{fig:qared} and $q=2$ in \fig~\ref{fig:qbred} is that
in the latter case there is a twofold degeneracy below the separatrix surface.
\def\figbcqb{%
Graphical evaluation of \equ~(\ref{eq:qcondition}) for two coupled rotors
with $\m = (-1,1)$, $q = 2$, $\eps=0.02$, $\hbar = 0.02$, and $\nb_1/2$
even. The dashed line marks the separatrix.}
\def\FIGbcqb{\centerline{\psfig{figure=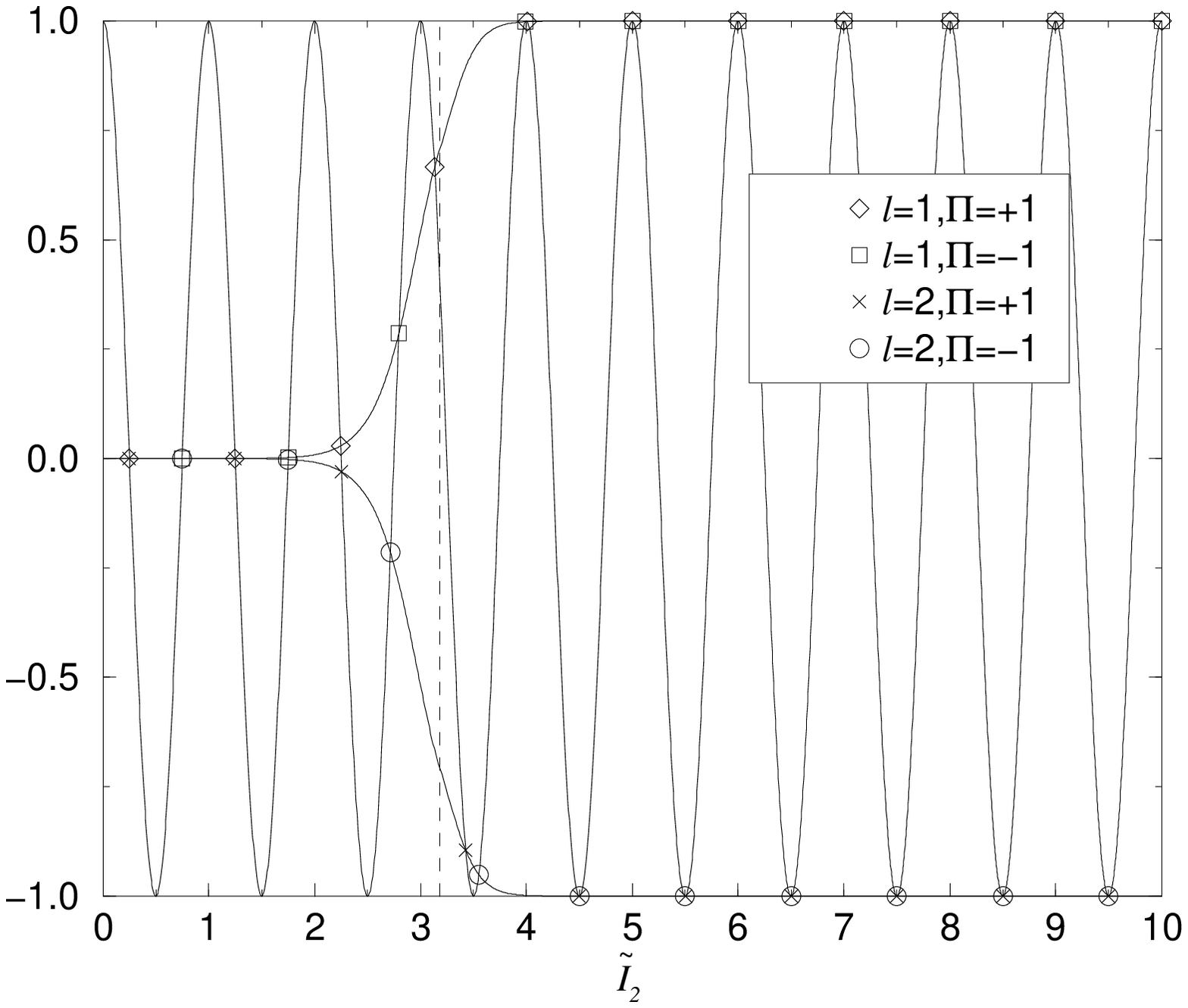,width=10.0cm}}}
\FIGo{fig:bcqb}{\figbcqb}{\FIGbcqb}
\def\figqbred{%
Semiclassical eigenvalues of two coupled rotors with $\m = (-1,1)$, $q = 2$,
$\eps = 0.02$, and $\hbar=0.02$ in $\BIcred$-space. The solid line is the
energy surface $\h=0.15$ and the dashed line is the separatrix
surface.}
\def\FIGqbred{\centerline{\psfig{figure=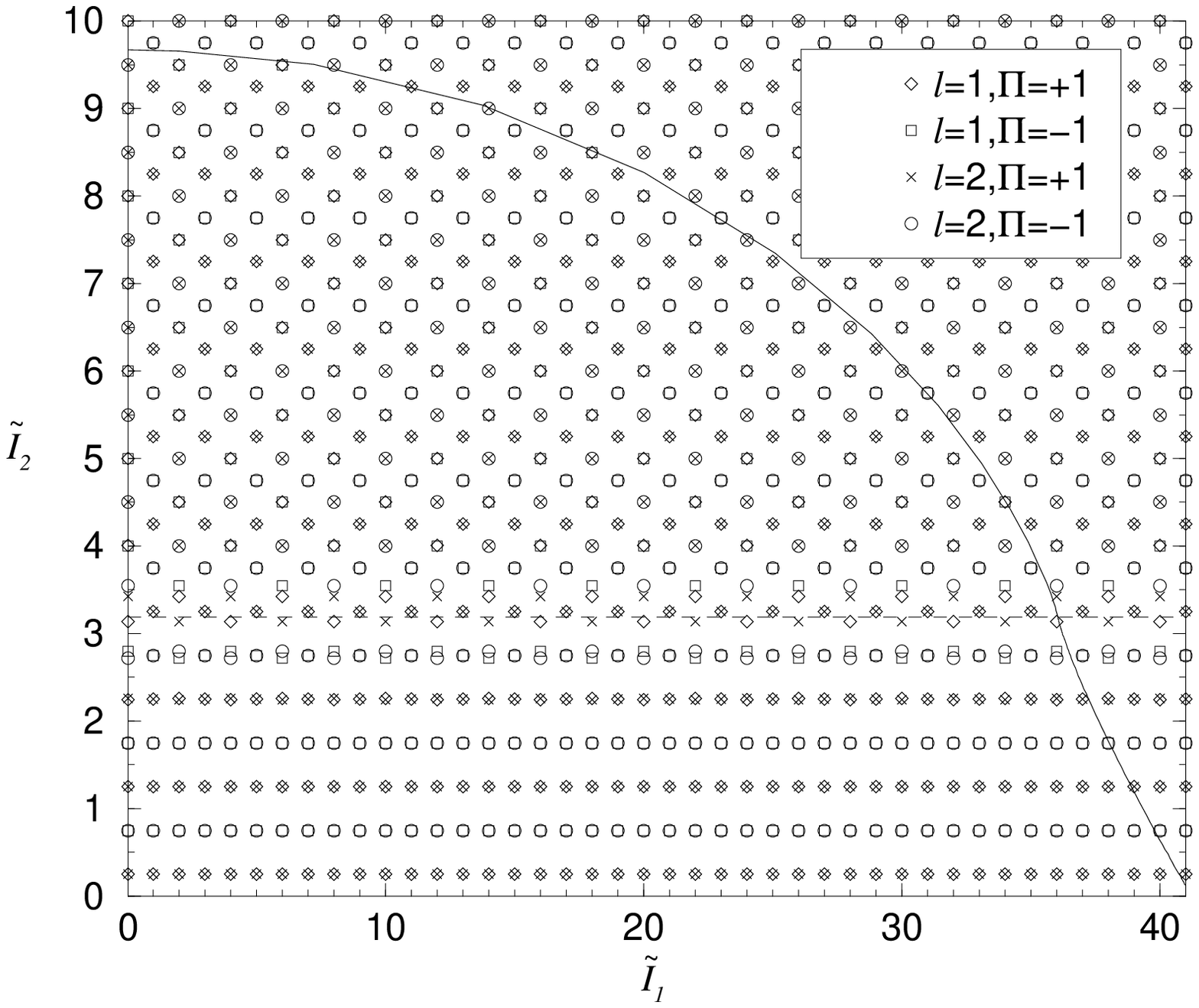,width=11.0cm}
}}
\FIGo{fig:qbred}{\figqbred}{\FIGqbred}

Finally, we illustrate a more generic case with irrational $\sigma$ using 
two coupled non-identical rotors as example
\begin{equation}
  H_0 = \frac{1}{2}\Ia_1^2+\frac{\gamma}{2}\Ia_2^2  \ ,
\end{equation} 
with $\gamma$ being the reciprocal of the golden mean, $(\sqrt{5}-1)/2$. 
We again take $\m = (-1,1)$ and $q=1$ giving $\sigma = -\nb_1/(1+\gamma)$. 
Figure~\ref{fig:qaredgm} confirms that the quadratic lattices below the
separatrix surface are as in the case of integer-valued $2\sigma$ pictured in
\fig~\ref{fig:qared}. But above the separatrix surface, there are two
non-congruent skewed lattices with basis vectors ${\bm a}_0 =
(0,\bar{\Pi}\hbar)$ and lattice vectors ${\bm a}_1 =
(\hbar,\bar{\Pi}\hbar/(1+\gamma))$, ${\bm a}_2 = (0,\hbar)$. A quantum cell
does not exist in this case, which is expressed by the seeming irregularity
of the overlap of the lattices in \fig~\ref{fig:qaredgm}. 
Note that the eigenvalues in \fig~\ref{fig:qaredgm} are
distinguished by their quantum number $\Pi$ and not by the lattice index
$\bar{\Pi}$.   
\def\figqaredgm{%
Semiclassical eigenvalues of two coupled non-identical rotors with irrational
$\sigma$, $\m = (-1,1)$, $q = 1$, $\eps = 0.02$, and $\hbar=0.02$ in
$\BIcred$-space. The dashed line marks the separatrix surface.
Filled regions represent elementary cells belonging to $\bar{\Pi} = +1$ (left)
and $\bar{\Pi} = -1$ (right).}
\def\FIGqaredgm{\centerline{\psfig{figure=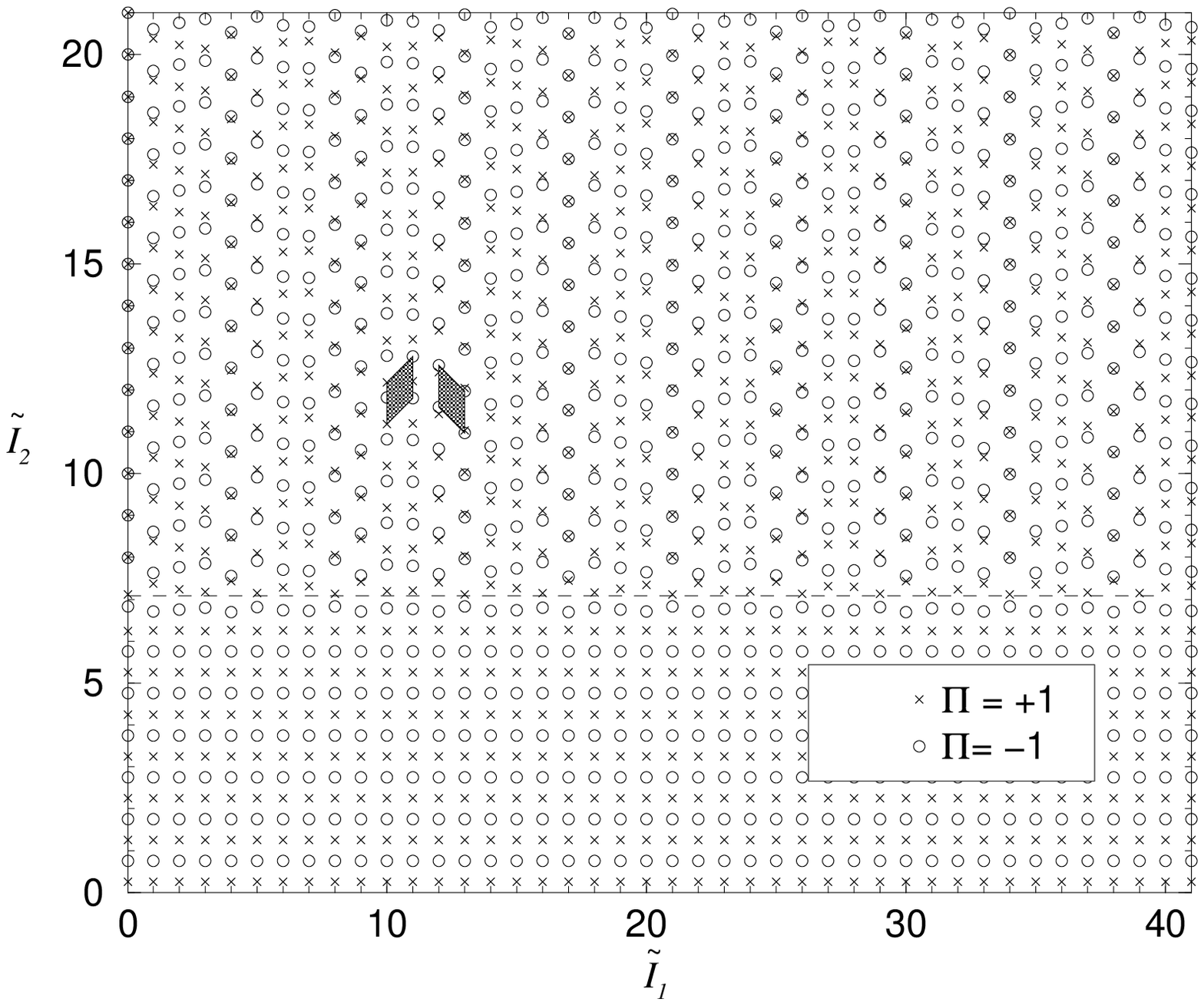,width=11.0cm}
}}
\FIGo{fig:qaredgm}{\figqaredgm}{\FIGqaredgm}

\section{CONCLUSIONS}
\label{sec:con}
The classical and quantum mechanics of isolated, nonlinear resonances has been
presented in action space.
The energy surfaces were found to be typically composed of two
different kinds of patches corresponding to the inner and the outer part of
the resonance zone. 
The graphical representation of these surfaces for a model of coupled
rotors proved to be a very concise description of the integrable dynamics.  
Moreover, it was demonstrated that energy surfaces outside the resonance zone
can be approximated by canonical perturbation theory, but have to be handled
with care, since non-physical segments are produced close to the resonance
zone.     

Exploiting the one-component property of the system, we have investigated the 
distribution of the quantum mechanical eigenvalues in the action space of the
symmetry reduced system: the eigenvalues within the resonance zone are located 
on $\dof$-dimensional cubic lattices, whereas the other eigenvalues are
located on locally $\dof$-dimensional triclinic lattices reflecting the
non-trivial symmetry reduction. Both kinds of lattices and the smooth
transition between them were described by a uniform semiclassical quantization
procedure and graphically illustrated with the help of the coupled-rotor
model.   
It was found that simple cubic lattices can be recovered only separately for
each of the different types of classical motion in a properly reduced action
space of the full system.    

The presented discussion deals with nonlinear resonances. The
important linear case of coupled harmonic oscillators demands special
considerations in a future publication.

\begin{acknowledgment}
I wish to thank P.H.~Richter for attracting my attention to this subject and
his support during my PhD studies.  
H.~Waalkens and O.~Zaitsev are acknowledged for critically reading the
manuscript.   
\end{acknowledgment}


\bibliographystyle{plain}
\bibliography{}

\rem{
\newpage
\section*{Figure captions}
\setlength{\parindent}{0cm}
\printfigcap{fig:portrait}{\figportrait}
\printfigcap{fig:esurfa}{\figesurfa}
\printfigcap{fig:esurfb}{\figesurfb}
\printfigcap{fig:wind}{\figwind}
\printfigcap{fig:foli}{\figfoli}
\printfigcap{fig:esurfbred}{\figesurfbred}
\printfigcap{fig:d2m23}{\figd2m23}
\printfigcap{fig:resa}{\figresa}
\printfigcap{fig:resb}{\figresb}
\printfigcap{fig:resc}{\figresc}
\printfigcap{fig:resd}{\figresd}
\printfigcap{fig:stef}{\figstef}
\printfigcap{fig:bcqa}{\figbcqa}
\printfigcap{fig:Xqa}{\figXqa}
\printfigcap{fig:qared}{\figqared}
\printfigcap{fig:qa}{\figqa}
\printfigcap{fig:bcqb}{\figbcqb}
\printfigcap{fig:qbred}{\figqbred}
\printfigcap{fig:qaredgm}{\figqaredgm}

\newpage
\setcounter{figure}{0}
\showfig{\FIGportrait}
\showfig{\FIGesurfa}
\showfig{\FIGesurfb}
\showfig{\FIGwind}
\showfig{\FIGfoli}
\showfig{\FIGesurfbred}
\showfig{\FIGd2m23}
\showfig{\FIGresa}
\showfig{\FIGresb}
\showfig{\FIGresc}
\showfig{\FIGresd}
\showfig{\FIGstef}
\showfig{\FIGbcqa}
\showfig{\FIGXqa}
\showfig{\FIGqared}
\showfig{\FIGqa}
\showfig{\FIGbcqb}
\showfig{\FIGqbred}
\showfig{\FIGqaredgm}
}

\end{article}
\end{document}